\documentclass[twocolumn,tighten,twocolappendix]{aastex63}

\usepackage{color, hyperref}
\usepackage[normalem]{ulem}
\usepackage{amsmath}
\usepackage{CJK}

\newcommand{\sgr}{Sagittarius}

\newcommand{\transpose}[1]{{#1}^{\!\mathsf T}}

\newcommand{\feh}{[\mathrm{Fe}/\mathrm{H}]}

\newcommand{\kms}{\mathrm{km} \, \mathrm{s}^{-1}}
\newcommand{\Ly}{L_{\rm y}}
\newcommand{\Lz}{L_{\rm z}}
\newcommand{\Vgsr}{V_{\rm GSR}}
\newcommand{\lunits}{{\rm kpc}\,{\rm km} \, {\rm s}^{-1}}
\newcommand{\Msun}{\,M_{\odot}}

\newcommand{\NbLerr}{79}     
\newcommand{\Ngiants}{6830} 
\newcommand{\NhiLx}{18}     
\newcommand{\Nbchem}{5}      
\newcommand{\Nsgr}{823}     

\usepackage{graphicx}
\DeclareGraphicsExtensions{.pdf,.pdf}
\graphicspath{{./}{figures/}}

\newcommand{\minesweeper}{\texttt{Minesweeper}}

\shorttitle{Sagittarius in H3}
\shortauthors{Johnson et al.}
\begin{document}
\begin{CJK*}{UTF8}{gbsn}

\title{A Diffuse Metal-Poor Component of the Sagittarius Stream Revealed by the H3 Survey}

\author[0000-0002-9280-7594]{Benjamin~D.~Johnson}
\affiliation{Center for Astrophysics $|$  Harvard \& Smithsonian, 60 Garden Street, Cambridge, MA 02138, USA}
\author[0000-0002-1590-8551]{Charlie~Conroy}
\affiliation{Center for Astrophysics $|$ Harvard \& Smithsonian, 60 Garden Street, Cambridge, MA 02138, USA}
\author[0000-0003-3997-5705]{Rohan~P.~Naidu}
\affiliation{Center for Astrophysics $|$  Harvard \& Smithsonian, 60 Garden Street, Cambridge, MA 02138, USA}
\author[0000-0002-7846-9787]{Ana~Bonaca}
\affiliation{Center for Astrophysics $|$  Harvard \& Smithsonian, 60 Garden Street, Cambridge, MA 02138, USA}
\author[0000-0002-5177-727X]{Dennis~Zaritsky}
\affiliation{Steward Observatory and University of Arizona, 933 N. Cherry Ave, Tucson AZ 85719}
\author[0000-0001-5082-9536]{Yuan-Sen~Ting(丁源森)}
\affiliation{Institute for Advanced Study, Princeton, NJ 08540, USA}
\affiliation{Department of Astrophysical Sciences, Princeton University, Princeton, NJ 08544, USA}
\affiliation{Observatories of the Carnegie Institution of Washington, 813 Santa Barbara Street, Pasadena, CA 91101, USA}
\affiliation{Research School of Astronomy and Astrophysics, Mount Stromlo Observatory, Cotter Road, Weston Creek, ACT 2611, Canberra, Australia}
\author[0000-0002-1617-8917]{Phillip~A.~Cargile}
\affiliation{Center for Astrophysics $|$  Harvard \& Smithsonian, 60 Garden Street, Cambridge, MA 02138, USA}
\author[0000-0002-6800-5778]{Jiwon~Jesse~Han}
\affiliation{Center for Astrophysics $|$  Harvard \& Smithsonian, 60 Garden Street, Cambridge, MA 02138, USA}
\author[0000-0003-2573-9832]{Joshua~S.~Speagle}
\affiliation{Center for Astrophysics $|$  Harvard \& Smithsonian, 60 Garden Street, Cambridge, MA 02138, USA}


\begin{abstract}

    The tidal disruption of the Sagittarius dwarf galaxy has generated a spectacular stream of stars wrapping around the entire Galaxy.  We use data from \emph{Gaia} and the H3 Stellar Spectroscopic Survey to identify $\Nsgr$ high-quality Sagittarius members based on their angular momenta. The H3 Survey is largely unbiased in metallicity, and so our sample of Sagittarius members is similarly unbiased.  Stream stars span a wide range in [Fe/H] from $-0.2$ to $\approx-3.0$, with a mean overall metallicity of $\langle$[Fe/H]$\rangle=-0.99$.  We identify a strong metallicity-dependence to the kinematics of the stream members.  At [Fe/H]$>-0.8$ nearly all members belong to the well-known cold ($\sigma_v< 20\,\kms$) leading and trailing arms.  At intermediate metallicities ($-1.9<$[Fe/H]$<-0.8$) a significant population (24\%) emerges of stars that are kinematically offset from the cold arms.  These stars also appear to have hotter kinematics.  At the lowest metallicities ([Fe/H]$\lesssim-2$), the majority of stars (69\%) belong to this kinematically-offset diffuse population.  Comparison to simulations suggests that the diffuse component was stripped from the Sagittarius progenitor at earlier epochs, and therefore resided at larger radius on average, compared to the colder metal-rich component.  We speculate that this kinematically diffuse, low metallicity, population is the stellar halo of the Sagittarius progenitor system.

\end{abstract}


\section{Introduction}

Streams of stars from tidally disrupted satellite galaxies provide insight into the build-up of the Milky Way through minor mergers and can be used as tracers of the Milky Way potential.  Accreted satellites likely contribute substantial numbers of stars to the Milky Way halo \citep[e.g.,][]{Bullock05, Bell08, Zolotov09, Cooper10, Monachesi19}, and may also influence the dynamical state \citep[e.g.,][]{Quinn86, Quinn93, Velazquez99, Font01, Kazantzidis08, Purcell11, Laporte18} and star formation history \citep[e.g.,][]{Hernquist95, Moreno15, Ruiz-Lara20} of the Milky Way disk.  Perhaps the most striking example of these processes is the ongoing tidal disruption of the \sgr{} dwarf galaxy.

\sgr{} was discovered as an overdensity of stars in velocity and position on the sky \citep{Ibata94}. Iso-density contours of a corresponding excess of stars at $R\sim18$ indicated that \sgr{} was highly elongated \citep{Ibata95}.  The use of luminous standard candles including RR Lyrae and M giant stars, and matched color magnitude diagram filtering allowed this elongation to be mapped in excess number counts to ever larger separations from the \sgr{} dwarf remnant, eventually reaching across the entire sky \citep[e.g.,][]{Mateo96, Alard96, Alcock97, Totten98, Mateo98, Majewski99, Ibata01, Newberg03, Majewski03, Belokurov06, Belokurov14, Sesar17, Hernitschek17}.  At the same time, spectroscopic followup was used to identify members of the prominent leading and trailing arms as coherent velocity overdensities often well-separated from the bulk of Milky Way stars \citep[e.g.,][]{Ibata97, Majewski99, Majewski04, Belokurov14}.  Detailed studies of the stellar population of \sgr{} have taken advantage of these overdensities to identify \sgr{} members via selections in position and line-of-sight velocity.

Exploration of the chemical composition of such \sgr{} members revealed a substantial metallicity difference between the dwarf galaxy remnant ([Fe/H]$\sim -0.4$) and the streams ([Fe/H] $\sim -1$) and suggestions of a gradient along the streams themselves \citep[e.g.][]{Bellazzini06, Chou07, Monaco07, Carlin12, Gibbons17}.  These observations were interpreted to suggest a steep metallicity gradient within the \sgr{} progenitor, and have implications for the composition of stars contributed by \sgr{} to the Milky Way halo.

With the release of \emph{Gaia} DR2 the proper motions and parallaxes of 2 billion stars across the entire sky became available \citep{GaiaDR2}.  While \sgr{} debris is too distant for significant detection of parallax by \emph{Gaia}, the proper motions of the luminous giants are of sufficient S/N to allow a clean separation against the background \citep{Antoja20}.  Moreover, supplementing the \emph{Gaia} data with external distances, e.g., from standard candles or CMD fitting, enables a more robust identification of \sgr{} debris \citep[e.g.,][]{Ibata20, Ramos20}.  However, these techniques lack the full 6D phase space information, and so contamination is still a source of concern.

The combination of \emph{Gaia} data with large spectroscopic surveys -- e.g., LAMOST, SEGUE, and APOGEE -- has provided a 6D phase space view of \sgr{} \citep{Li19, Yang19, Hayes20}.  This has allowed for selection of \sgr{} members based on conserved quantities such as integrals of motion \citep{Li19, Yang19, Hayes20}, as well as further insight into the metallicity distribution function (MDF) of \sgr{} stars and its variation along the streams. However, existing spectroscopic surveys are limited by small numbers and/or significant selection biases in metallicity.

In this work we identify \sgr{} members in the H3 Stellar Spectroscopic Survey on the basis of their Galactocentric angular momentum.  A key feature of the H3 Survey is that the main sample is selected solely on the basis of apparent magnitudes and {\it Gaia} parallaxes, so the resulting sample is largely unbiased with respect to metallicity.  In Section \ref{sec:data} we describe the H3 Survey and how we correct for the selection function.  In Section \ref{sec:simulations} we describe two simulations of the \sgr{} system that are used to inform our selection of \sgr{} members and our interpretation of the data.  In Section \ref{sec:results} we detail the selection of \sgr{} members in the H3 survey and use the identified stars to explore the metallicity distribution function (MDF) and kinematics of \sgr{}.  In Section \ref{s:comp2models} we compare our data to simulations of the \sgr{} tidal streams with a focus on possible correspondences to this low metallicity, diffuse population.  Finally, in Section \ref{sec:discussion} we discuss the possible origins of this population and the implications of a low metallicity stellar halo of the \sgr{} progenitor.

\section{Data}
\label{sec:data}

\subsection{Overview and Derived Quantities}

In this paper we combine data from the {\it Gaia} satellite and the H3 Survey \citep{Conroy19a}.  {\it Gaia} is delivering parallaxes and proper motions for $>1$ billion stars to $G\approx20$.  H3 is a medium resolution ($R\approx32,000$) spectroscopic survey of stars in the northern hemisphere and at high Galactic latitudes.  Specifically, the primary H3 selection function is $|b|>30^\circ$, $15<r<18$, and $\pi<0.5$ mas, where the latter is a selection on the {\it Gaia} parallax. To date, all but a handful of the currently acquired fields are at $|b|>40^{\circ}$. The selection enables efficient targeting of distant halo stars, which is the primary scientific motivation of the survey.  Critically, the main H3 selection function is largely unbiased with respect to metallicity, as no color-cuts are applied.  The H3 survey also includes an additional secondary selection of rare and distant K giants and blue horizontal branch (BHB) stars \citep{Conroy19a}, which we include in this work with appropriate re-weighting where necessary (see Section \ref{sec:reweighting} below).

Stellar parameters and distances are derived for each star using \minesweeper{}  \citep{Cargile19}.  Briefly, \minesweeper{} is a Bayesian inference program that fits the combined H3 spectrum and broadband photometry to a library of stellar isochrones and synthetic spectral models.  For most of the H3 sample the {\it Gaia} parallax is low S/N (as the stars are distant and hence have small parallaxes).  The {\it Gaia} parallax is included as a prior in the fitting; this prior is helpful for separating dwarfs and giants even when the parallax S/N is low.  The fit parameters include the radial velocity, stellar mass, age, [Fe/H], [$\alpha$/Fe], $A_{\rm V}$, and heliocentric distance.  \citet{Cargile19} validate this approach using a variety of mock data, star cluster data, high-quality benchmark stars, duplicate H3 observations, and a subset of the H3 data that has high S/N {\it Gaia} parallaxes.  These tests demonstrate that the H3 pipeline is delivering reliable stellar parameters, with systematic uncertainties in radial velocities of $\lesssim 1\,\kms$ and metallicities of $\lesssim 0.1$ dex.

From the basic 6D phase space quantities of radial velocity, distance, R.A., Dec., and proper motions, we compute a wide array of derived quantities including angular momenta and orbital energies.  The latter require adopting a Galactic potential.  Here we use the default \texttt{MilkyWayPotential} in \texttt{gala v1.1} \citep{gala1, gala2, bovy15}.  We adopt the Galactocentric frame \texttt{v4.0} implemented in \texttt{Astropy v4.0} \citep{astropy1, astropy2}.  We also compute the radial velocity projected into the Galactocentric Standard of Rest (GSR), $V_{\rm GSR}$.  It is important to note that this quantity is distance-independent --- it is a function of only radial velocity and sky coordinates.  Heliocentric \sgr{} stream coordinates are computed using the frame of \citet{Majewski03} as implemented in \texttt{gala}; the stream longitude coordinate $\Lambda_{\rm Sgr}$ increases from the progenitor towards the trailing stream. Uncertainties on these quantities are propagated using the the posterior samples obtained from \minesweeper{} for distance and radial velocity along with assumed Gaussian uncertainties for {\it Gaia} proper motions.

We use the H3 catalog V2.4, which contains 125,000 stars observed through Feb, 2020. Here we focus on stars with $\log g < 3.5$ to remove the dwarf stars that are at much smaller distances than the bulk of \sgr{}. In addition we require a median spectroscopic S/N $> 3$, and remove stars flagged for known issues in the data analysis.  We also remove $\NbLerr$ stars with large uncertainties in their angular momenta ($> 3\times 10^3 \,\, \lunits$).  This results in a sample of $\Ngiants$ giants.

\subsection{Selection Function Re-Weighting}
\label{sec:reweighting}

Any survey provides an incomplete view of the sky, whether because of the survey geometry (window function), magnitude limit, and/or other selections (e.g., color-cuts).  For example, a magnitude-limited survey will be biased toward more nearby stars, and so a histogram of stellar metallicities from such a sample will be weighted toward the more nearby stars.  In order to provide a more complete view, one can re-weight the existing stars to account for the survey selection function, or forward model the entire process with a detailed model of the underlying population(s) \citep[see e.g.,][]{Rix13}.  We describe in this section our approach to re-weighting stars in the H3 Survey.

The primary H3 target selection does not impose an explicit metallicity bias (e.g., due to cuts in color space).  However, there is an additional color selection of K giants and BHB stars \citep{Conroy19a} that are assigned higher priority ranking in fiber assignment (these stars are rare, accounting for only $\sim 1-2$ stars per field).  Moreover the magnitude limit imparts a large distance bias, as well as a small distance-dependent metallicity bias.  To account for these effects we estimate the number of stars of a certain stellar type and priority ranking $p$ and above a given S/N threshold that are contributed to the catalog by pointing $i$ as
\begin{equation}
\lambda_i  = \Omega \, d^2  \, n_i({\rm [Fe/H]}, d) \, f_{\rm t} \, f_{i, {\rm m}} \, f_{i, {\rm p}} \, {\rm d} d,
\end{equation}
\noindent
where $\Omega$ is the solid angle of the pointing, $n$ is the density of all stars in the direction of the pointing at a particular heliocentric distance $d$ and metallicity ${\rm [Fe/H]}$, $f_{\rm t}$ is the fraction of stars that are of the selected stellar type (e.g., $\log g < 3.5$), $f_{i, {\rm m}}$ is the fraction of stars of that type that fall within a magnitude range that would be observable above the S/N threshold, and $f_{i, {\rm p}}$ is the fraction of stars of that priority rank that are assigned a fiber (which is independent of distance, metallicity, or magnitude.)

It is not our intention here to construct and fit a detailed model for the spatial and metallicity variation of $n$, accounting for Poisson statistics. However, we can estimate the effect of the selection function on the metallicity distribution with the following approach.  We assume that every star in the catalog represents $w \equiv K / (f_{\rm t} \, f_{i, {\rm m}} \, f_{i, {\rm p}})$ stars where $K$ is a normalizing constant, and then re-weight each star to account for these metallicity, distance, and priority class dependent selection effects.   Using a relationship between magnitude and S/N determined for each pointing along with foreground reddening, the product $(f_{\rm t} \, f_{i, {\rm m}})$ is computed for each star from isochrones, and is distance and metallicity dependent.  The factor $f_{i, {\rm p}}$ is computed from information about which available sources were assigned to fibers.  We apply these weights when constructing the overall MDF of \sgr{} in \S\ref{sec:results} below and find little difference from the MDF in raw number counts.


\begin{figure*}
\includegraphics[width=\textwidth]{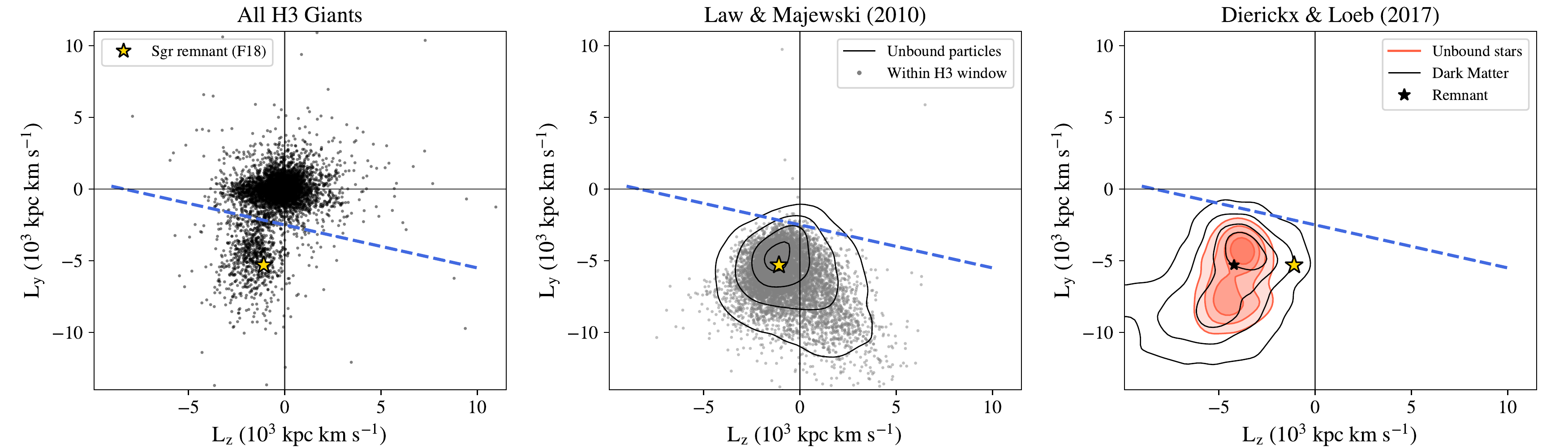}
\caption{Distribution in the $\Lz-\Ly$ plane of H3 giants (left) and particles from the LM10 (middle) and DL17 (right) simulations. The dashed blue lines show the selection criteria for identifying \sgr{} stars in H3 data.  Note that this selection naturally separates the two visible clumps in the data. In all three panels the observed angular momentum of the \sgr{} remnant is shown as a yellow star \citep[][F18]{Fritz18}.  In the middle panel, the contours correspond to all unbound LM10 stars, while the grey points correspond to the stars within the H3 window and magnitude selection function, with realistic noise applied to their distances and proper-motions.  In the right panel, all unbound stars and dark matter particles are shown as orange and black contours, respectively.  The simulated remnant is shown as a black star.
\label{fig:lylz}}
\end{figure*}

\section{Simulations of the Sagittarius Stream}
\label{sec:simulations}

To guide our interpretation of the H3 data we consider two N-body simulations of the tidal disruption of the \sgr{} dwarf galaxy in the halo of the Milky Way.

The first is the landmark simulation of \citet[][LM10 hereafter]{LM10}.  This simulation integrated the trajectories of particles representing the \sgr{} dwarf galaxy within a static Milky Way potential over 8 Gyr.  All the \sgr{} particles were initially distributed as a Plummer sphere \citep{Plummer1911}, and no distinction was made between stellar and dark matter particles. This simulation was able to obtain a very good match to the existing observational constraints on the \sgr{} tidal streams by varying the parameters of the Milky Way potential and the mass of the \sgr{} progenitor.  A triaxial potential was required to simultaneously produce the positions and radial velocities of stars in the leading stream. The initial mass and scale radius of the progenitor in the best-fitting simulation were $6.4\times 10^8 \, {\rm M}_\odot$ and $0.85 \, {\rm kpc}$.  In addition to 6D phase space information for each particle, LM10 provide the time when the particle became unbound from the \sgr{} progenitor and the rank-order of the energy of the particle within the progenitor (see LM10 for details). We have used the latter to compute $\hat{\rm R}_{\rm prog}$, the mean internal orbital radius of each particle within the Plummer potential of the progenitor; particles that are more tightly bound are more typically found in the inner regions of the \sgr{} progenitor. The time when the particle became unbound is tightly although non-linearly correlated with angular distance from the remnant, but only roughly correlated with $\hat{\rm R}_{\rm prog}$ since during pericentric passage particles with a large range of mean orbital radii can be stripped from the progenitor.

More recent simulations of the disruption of the \sgr{} progenitor have sought to explain new observations, especially of the distant trailing stream apocenter, and also to include more physical effects than LM10, such as a dynamic Milky Way halo \citep[e.g.,][]{Gibbons14, DL17, Laporte18, Fardal19}.  Here we consider the simulations of \citet[][DL17 hereafter]{DL17} due to the more massive and complex progenitor, featuring a dark, extended Hernquist halo with $M=1.3\times 10^{10} \,{\rm M}_\odot$ as well as stellar components in a more compact bulge and disk with 10\% of the halo mass.  The DL17 simulation tracked the orbits of both stellar and dark matter particles within a live Milky Way halo, thus accounting for a time-varying potential and dynamical friction. The simulation tracks the infall of \sgr{} from well beyond the Milky Way virial radius.  While effort was made to match the observed relative positions and velocities of the Sun, Galactic center, and \sgr{} remnant, the resulting simulation has substantial differences in detail from the actual observed stream properties, manifest largely as a coherent shift of the stream from observed coordinates.  For our purposes, this simulation will prove useful to highlight the different behavior of diffuse, less strongly bound particles (the dark matter halo) from the more strongly bound stellar particles in the presence of dynamical friction.

For both simulations we use the observational phase-space quantities (R.A., Dec., heliocentric distance, proper motions, and radial velocity) as reported by the authors and, when necessary, convert these to Galactocentric coordinates using the respective reference frames of those authors.  In order to create H3-like mock catalogs we have identified simulation particles that fall within the H3 window function.  For the LM10 particles we also generate mock photometry and apply a magnitude limit of $15<r<18$.  We then use the mock photometry to assign proper-motion uncertainties appropriate for \emph{Gaia}. We assign 10\% uncertainties in distance to both the LM10 and DL17 mocks; the median formal distance uncertainty for the H3 \sgr{} sample discussed below is 7\%.  Where noted we perturb the mock values by these uncertainties to create more realistic comparisons for the H3 data.

\section{Results}
\label{sec:results}

\begin{figure}[!t]
\includegraphics[width=0.48\textwidth]{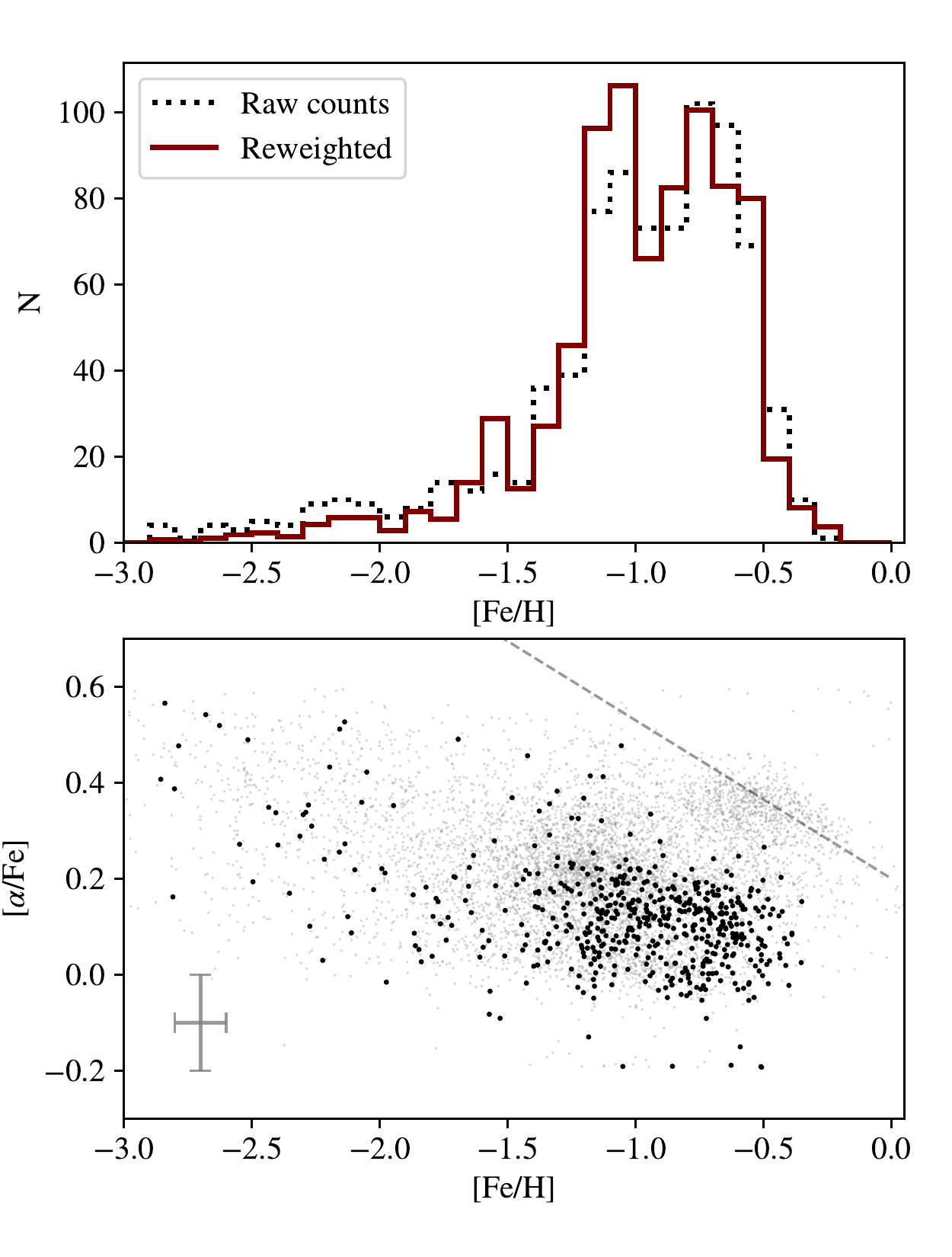}
\caption{Top panel: Metallicity distribution of \sgr{} members. Both the raw number counts and the re-weighted distribution (normalized to the same total number) are shown. Bottom panel: Distribution of \sgr{} members with S/N$>5$ in [Fe/H] and [$\alpha$/Fe] (black) and the entire H3 sample of S/N$>5$ giants (light grey).  A typical error bar is shown in the bottom left, and the definition of anomalous chemistry is shown as the dashed grey line.
\label{fig:mdf}}
\end{figure}

\begin{figure}
    \includegraphics[width=0.5\textwidth]{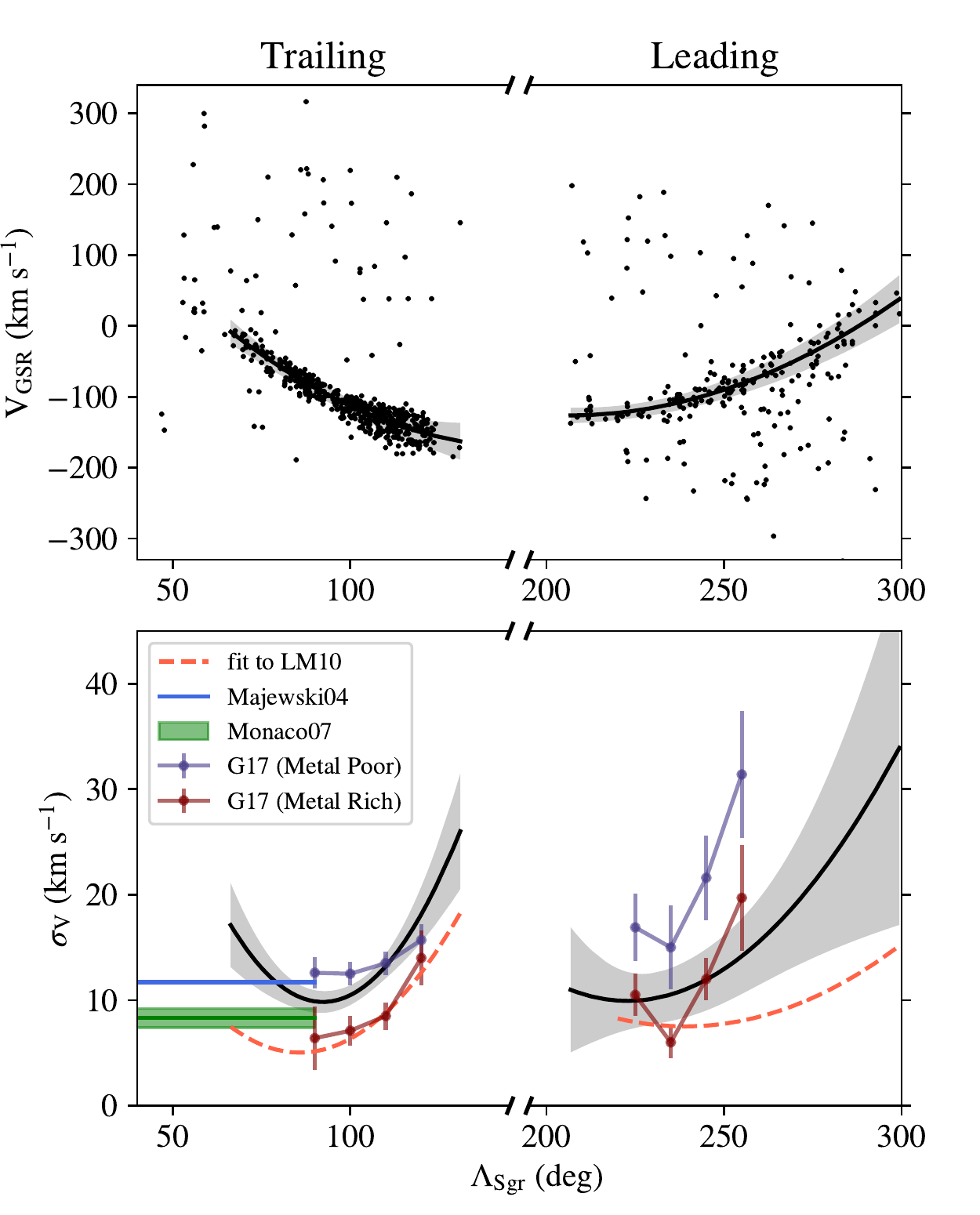}
    \caption{Top panel: $\Vgsr$ vs. stream longitude ($\Lambda_{\rm Sgr}$) for the H3 \sgr{} members that satisfy the angular momentum selection in Figure \ref{fig:lylz}.  Cold and diffuse components are clearly visible.  The highest posterior probability sample for the model for $\Vgsr(\Lambda_{\rm Sgr})$ in the cold component is shown as a solid line, while the corresponding dispersion is shown as a grey band.  Bottom panel: Velocity dispersion of the cold component as a function of stream longitude. Posterior median values and uncertainties from this work are shown in black/grey, and are compared to previous work \citep{Majewski04, Monaco07, Gibbons17}.  We also show our fit to the LM10 simulated stream as a dashed orange line.
    \label{fig:vel_fit}}
\end{figure}

\begin{deluxetable}{lrrrr}
    \tablecaption{Fitted Stream Parameters \label{tbl:fit}}
    \tablehead{
        \colhead{Parameter} & \colhead{MAP\tablenotemark{a}} & \colhead{Posterior} & \colhead{Prior Range}
        }
    \startdata
    \cutinhead{$\Lambda_{\rm Sgr} < 140^\circ$}
        $\alpha_0$ & $352$     & ${352}^{+26.8}_{-26}$            & $(250, 450)$ \\
        $\alpha_1$ & $-6.97$   & ${-6.96}^{+0.543}_{-0.557}$      & $(-10, 0)$ \\
        $\alpha_2$ & $0.023$  & ${0.023}^{+0.0029}_{-0.0028}$     & $(-0.05, +0.05)$ \\
        $\beta_0$  & $-102$    & ${-100}^{+23.9}_{-22.9}$         & $(-200, 0)$ \\
        $\beta_1$  & $2$       & ${1.96}^{+0.477}_{-0.492}$       & $(-1, 4)$ \\
        $\beta_2$  & $-0.011$ & ${-0.011}^{+0.0025}_{-0.0025}$    & $(-0.05, +0.05)$ \\
        $\mu_b$    & $90.2$    & ${83.7}^{+22.6}_{-23}$           & $(-150, +150)$ \\
        $\sigma_b$ & $158$     & ${160}^{+17.3}_{-14.5}$          & $(100, 250)$ \\
        $f$        & $0.094$  & ${0.097}^{+0.014}_{-0.012}$       & $(0, 0.3)$ \\
    \cutinhead{$\Lambda_{\rm Sgr} >  200^\circ$}
        $\alpha_0$ & $676$     & ${666}^{+173}_{-169}$            & $(200, 1500)$ \\
        $\alpha_1$ & $-7.78$   & ${-7.72}^{+1.4}_{-1.41}$         & $(-20, 0)$ \\
        $\alpha_2$ & $0.0189$  & ${0.0188}^{+0.0029}_{-0.0029}$   & $(-0.05, 0.05)$ \\
        $\beta_0$  & $213$     & ${200}^{+152}_{-159}$            & $(-100, 500)$ \\
        $\beta_1$  & $-1.82$   & ${-1.73}^{+1.35}_{-1.25}$        & $(-5, 5)$ \\
        $\beta_2$  & $0.00409$ & ${0.0039}^{+0.0026}_{-0.0029}$   & $(-0.03, 0.03)$ \\
        $\mu_b$    & $-74.6$   & ${-81.5}^{+14.1}_{-14}$          & $(-150, 150)$ \\
        $\sigma_b$ & $138$     & ${139}^{+11.5}_{-9.86}$          & $(100, 250)$ \\
        $f$        & $0.41$    & ${0.408}^{+0.039}_{-0.038}$      & $(0, 0.6)$
    \enddata
    \tablenotetext{a}{Maximum {\it a posteriori} sample in the Monte Carlo chain.}
    \tablecomments{Reported posterior parameter values are the 50th percentile of the marginalized posterior PDFs, while uncertainties are computed from the 16th and 84th percentiles. There are strong covariances between parameters.}
\end{deluxetable}

\begin{figure*}[!t]
\includegraphics[width=\textwidth]{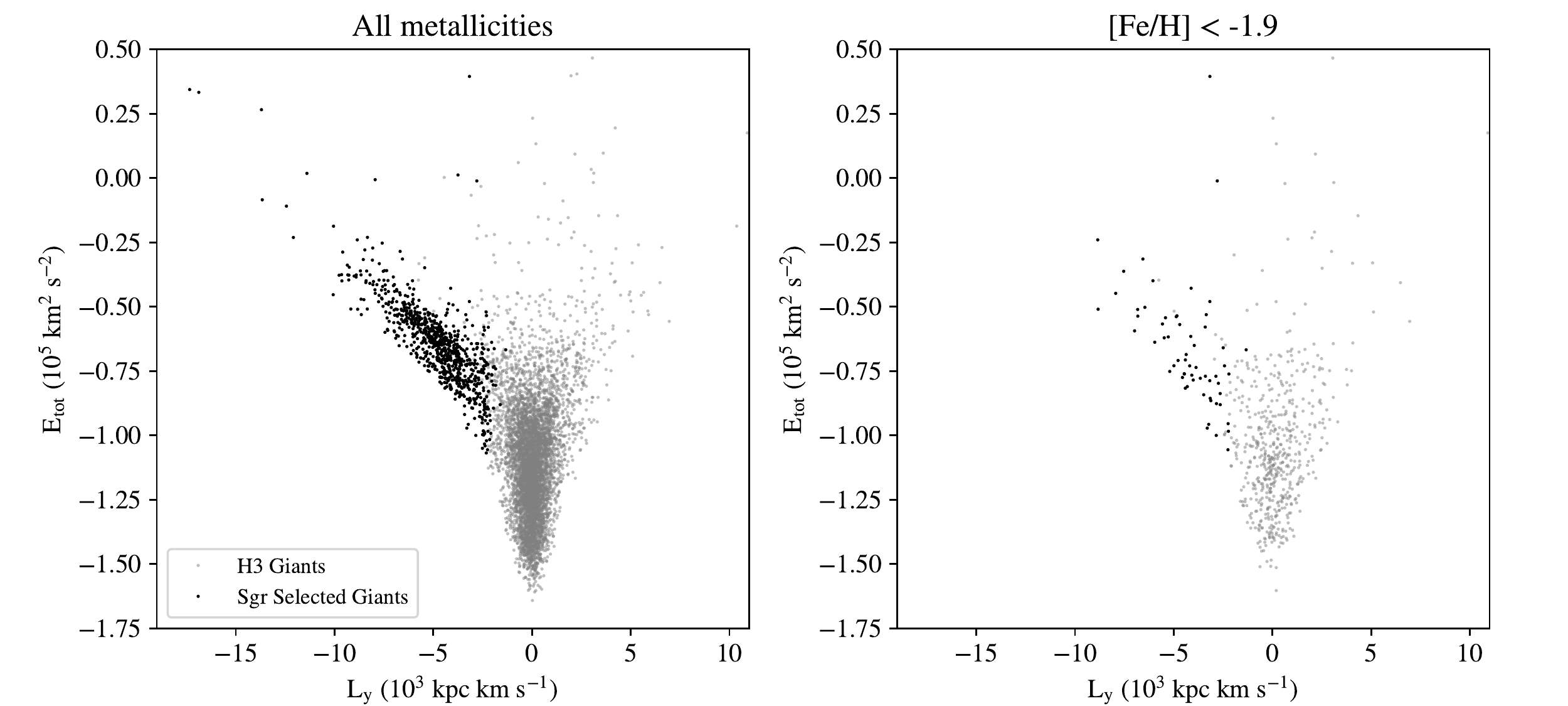}
\caption{Distribution of H3 giants in the space of orbital energy ($E_{\rm tot}$) and the $y-$component of angular momentum ($\Ly$).  \sgr{} members are shown in black.  The left panel shows all stars, while the right panel shows stars with [Fe/H]$<-1.9$.  \sgr{} members are clearly identified as the ``spur" extending toward negative $\Ly$.  The parallel diagonal tracks in the left panel are associated with the leading and trailing arms.
\label{fig:ely}}
\end{figure*}

\begin{figure*}
\includegraphics[width=\textwidth]{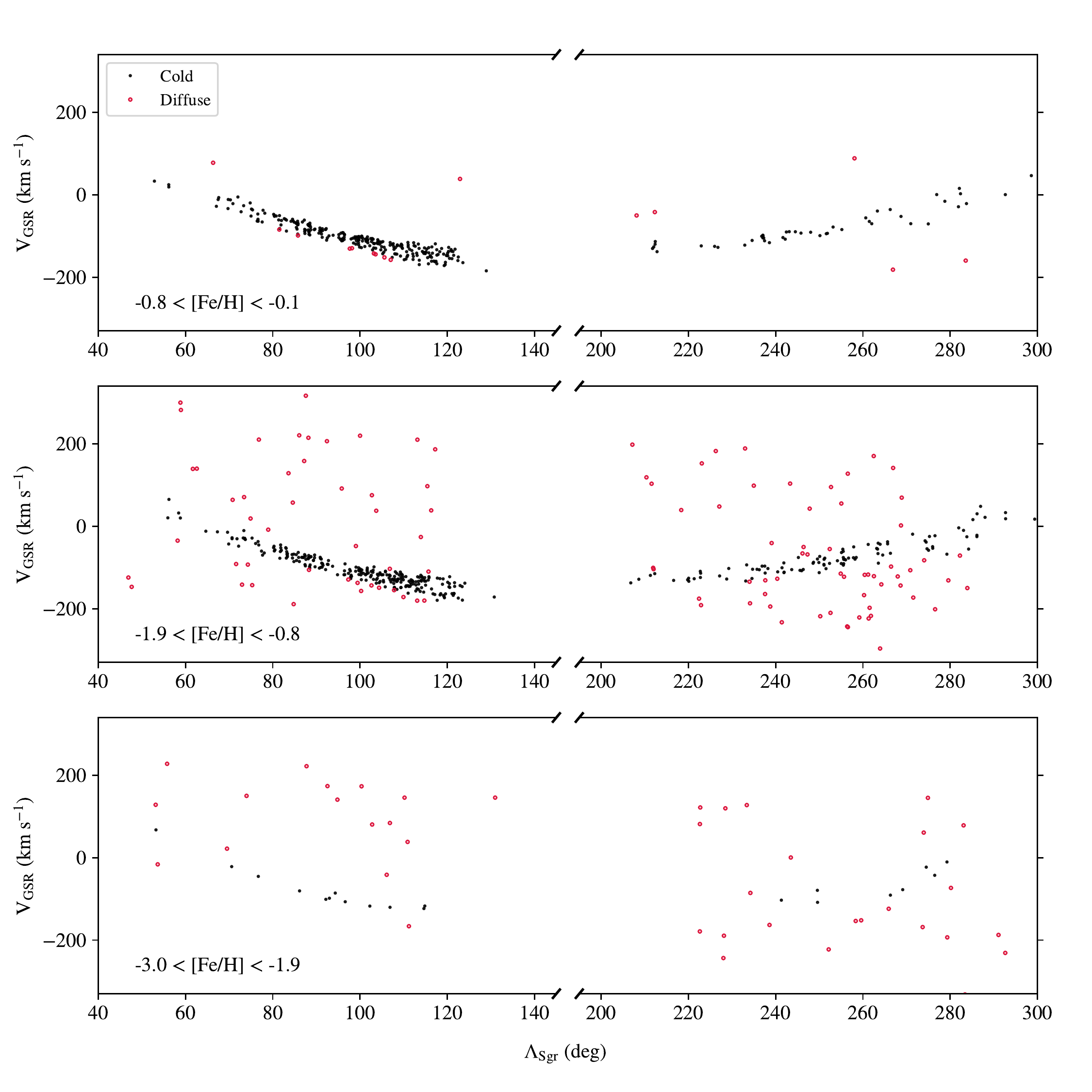}
\caption{
  $\Vgsr$ as a function of \sgr{} stream longitude for \sgr{} members in H3.  The sample is split into three metallicity bins, one for each panel.  Within each panel, stars are associated either with a kinematically cold or diffuse component (filled vs. open symbols, see Section \ref{s:vel_fit} for details).   Notice that the diffuse component is much more prominent at lower metallicities.  For the kinematically cold component, the stars at $\Lambda_{\rm Sgr}<140^\circ$ are associated with the trailing arm, while those at $\Lambda_{\rm Sgr}>200^\circ$ are associated with the leading arm.
\label{fig:vgsrfeh}}
\end{figure*}

\begin{figure}
\includegraphics[width=0.5\textwidth]{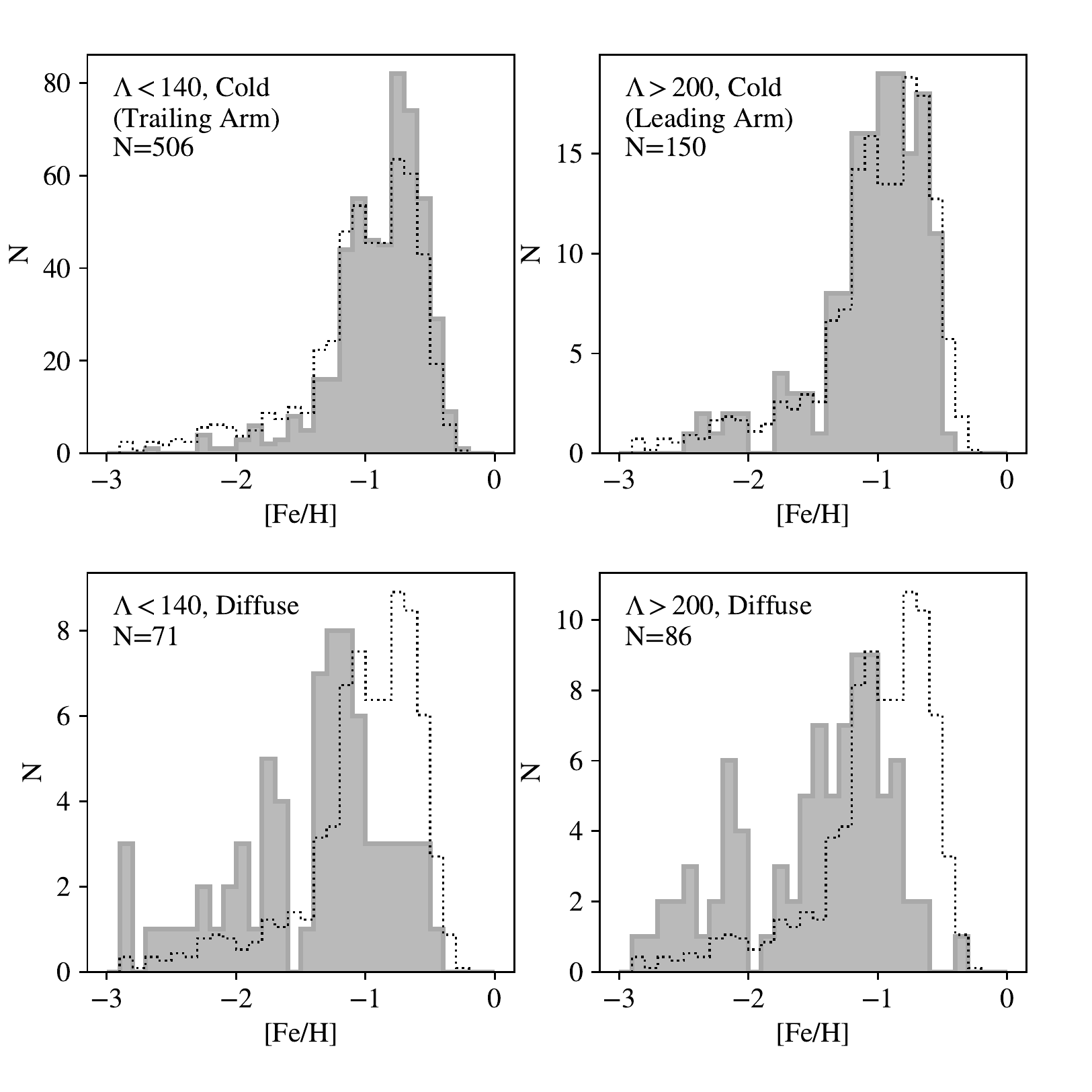}
\caption{Metallicity distribution function (MDF) in raw counts of \sgr{} members separated into trailing and leading arms (top left and right panels), and cold vs. diffuse components (top and bottom panels).  The overall unweighted MDF is shown as a dotted line, renormalized to the total number of stars in each component.  The leading cold arm is on average slightly more metal-poor than the trailing cold arm.  The diffuse components are more metal-poor on average than the cold components and display a greater fraction of very metal-poor stars.
\label{fig:mdfvel}}
\end{figure}

\subsection{Selection of Sagittarius Members}
\label{sec:selection}

The H3 Survey enables the measurement of 6D phase space coordinates for a homogeneously selected sample of distant stars.  The \sgr{} angular momentum vector is nearly aligned with the negative Galactic $y-$axis, which means that \sgr{} stars will have high values of $\Ly$ - angular momentum in the negative $y-$direction.  Except for the effects of dynamical friction and non-spherical potentials, the angular momenta of stars in the Milky Way halo are expected to remain a conserved quantity.

Figure \ref{fig:lylz} shows the $\Ly-\Lz$ plane for H3 giants and the LM10 and DL17 simulations.  As expected, the simulations are confined to negative values of $\Ly$ close the the modeled \sgr{} remnant, even in the presence of a non-spherical potential (LM10) or dynamical friction (DL17).  In the data there is also a clear excess of stars in a region of angular momentum space occupied by the simulations.  We therefore define a simple selection in this plane that encompasses the vast majority of the mock particles in the LM10 simulations and also separates the bimodal distribution of H3 giant stars.  Specifically, we select \sgr{} stars with
\begin{eqnarray}
\Ly < -2.5 - 0.3\, \Lz,
\end{eqnarray}
where the angular momenta are in units of $10^3 \,\, \lunits$. We remove from this selection $\NhiLx$ stars with a substantial fraction of their angular momentum in the $x$ direction and $\Nbchem$ stars with clearly anomalous chemistry ($[\alpha/\rm{Fe}] > -0.3 \, [\rm{Fe}/\rm{H}] + 0.2$).  These selections result in a sample of $\Nsgr$ \sgr{} stars, which is $\sim 12$\% of the giants in the H3 Survey.

While the \sgr{} stars form a prominent locus in the $\Ly-\Lz$ plane and our selection criterion runs through the minimum between this and the locus defining the bulk of the H3 giants, there may still be contamination by non-\sgr{} members.  This could arise from intrinsic overlap in the distribution of \sgr{} and non-\sgr{} stars in this plane, or from scattering of non-\sgr{} stars into our simple selection by angular momentum errors. Removal of a non-uniform ``background" population of stars contributed by distinct halo structures \citep[e.g.][]{Naidu20} is difficult; probabilistic classification and selection may prove useful in future work, as well as consideration of additional observables (e.g. chemistry).
Finally, we note that detailed investigation of the H3 giant sample has revealed a small number giant stars with inferred distances that are too small by a factor of approximately two due to confusion between the red clump and red giant branch.  Of these, $65$ have proper motions consistent with \sgr{} and fall within our selection criterion when their distances are doubled.  Future work to resolve this issue will add to the number of identified \sgr{} members in the H3 survey.

\subsection{Global Metallicities and Abundances}

We begin by considering the metallicity distribution function (MDF) for the \sgr{} tidal debris in H3.  In previous work on this topic \sgr{} members were identified via RR Lyrae, M giants, or other color-selected samples.  These allowed for efficient selection of members, but at the cost of imparting significant biases in the metallicities of the resulting sample \citep[see][for a discussion of some of these issues]{Conroy19b}.  H3 is unique in this regard, in that the selection of spectroscopic targets is largely unbiased with respect to metallicity.

In the top panel of Figure \ref{fig:mdf} we show the MDF for \sgr{} stream members in H3, excluding the 10 specially selected BHB stars due to their large metallicity uncertainties.  The MDF is shown both with and without the corrections described in Section \ref{sec:reweighting}, demonstrating that the survey selection function has little effect on the overall MDF. The stream is quite metal-rich, with a mean metallicity of $\langle\feh\rangle=-1.02$. The weighted mean accounting for selection biases is $\langle\feh\rangle=-0.99$. This is comparable to recent estimates from the APOGEE Survey \citep{Hayes20}.  We also see a significant tail of metal-poor stars, extending to $\feh\approx-3$.  This tail comprises 49 stars with $\feh<-2.0$, accounting for $\sim 6$\% of the \sgr{} sample in raw numbers or 3\% when re-weighting. This does not include the 10 BHB stars with [Fe/H]$\lesssim -2$, which would bring the fraction of metal poor stars to 7\% in raw numbers. The nature of these metal-poor stars will be explored in detail below.

In the bottom panel of Figure \ref{fig:mdf} we show the distribution of \sgr{} stream members with S/N$>5$ in [$\alpha$/Fe] vs. [Fe/H].  Overall, the \sgr{} members have relatively low [$\alpha$/Fe] abundances compared to the general population of giants in H3 \citep{Conroy19b}. Indeed, the \sgr{} [$\alpha$/Fe] abundances are the lowest of all halo components identified in the H3 survey \citep{Naidu20}. We note that our [$\alpha$/Fe] abundances are somewhat higher than measured from APOGEE data \citep{Hasselquist19, Hayes20}.

\subsection{Identifying Kinematically Cold and Diffuse Populations}
\label{s:vel_fit}

The trend of $\Vgsr$ as a function of \sgr{} stream longitude, $\Lambda_{\rm Sgr}$, has traditionally proven a powerful way to identify \sgr{} members and to constrain models of the \sgr{} stream.  This particular space is advantageous because neither quantity depends on distances or proper motions (which dominate the error budget), and because the leading and trailing arms of \sgr{} are visible as cold structures.  In this paper we focus on this space for similar reasons, and only use the distances and proper motions to select probable \sgr{} members in angular momentum space (see Figure \ref{fig:lylz}).

In the top panel of Figure \ref{fig:vel_fit} we show $\Vgsr$ as a function of stream longitude for the H3 stars selected as \sgr{} members by their angular momentum.  The previously known cold components of the leading ($\Lambda_{\rm Sgr}>200^\circ$) and trailing ($\Lambda_{\rm Sgr}<140^\circ$) arms are clearly apparent.  There is also a population of stars more broadly distributed in $\Vgsr$.  Without full 6D phase space information this population would have been relegated to a background; with full 6D phase space information we now know that they have angular momenta that clearly associate them with \sgr{}.

To decompose the \sgr{} stars into kinematically cold and diffuse components we model the run of mean $\Vgsr$ with \sgr{} stream longitude as two second order polynomials, one for the cold trailing stream and one for the cold leading stream.  We further model the dispersion of $\Vgsr$ in these components as second order polynomials of the stream longitude.  Finally, we include a diffuse component modeled for simplicity as a single broad Gaussian in each longitude range with free mean and dispersion.  The fraction of stars belonging to this diffuse component is also left as a free parameter.

The likelihood of the H3 data for this model is
\begin{align}
\mathcal{L}  =  \prod_i \frac{1-f}{\sigma_{v, i}\sqrt{2\pi}}\, e^{-\frac{(v_i-\mu_{v,i})^2}{2\sigma^2_{v, i}}} +
                        \frac{f}{\sigma_{b}\sqrt{2\pi}} \, e^{-\frac{(v_i-\mu_{b})^2}{2\sigma^2_{b}}} \\
\mu_{v,i} = \transpose{\boldsymbol{\alpha}} \, {\boldsymbol{\lambda}}_i \quad ,  \quad \sigma_{v,i} =\transpose{\boldsymbol{\beta}} \, {\boldsymbol{\lambda}}_i
\end{align}
where $\boldsymbol{\alpha}$ and $\boldsymbol{\beta}$ are the 3-element vectors giving the coefficients of the polynomials for mean velocity and velocity dispersion respectively,
$\boldsymbol{\lambda}_i$ is the Vandermonde matrix of stream longitude for star $i$,
$f$ is the diffuse fraction, and $\mu_b$ and $\sigma_b$ are the mean and dispersion of the diffuse component. We infer the parameters of this model through nested Monte Carlo sampling \citep{Skilling04, dynesty} of the posterior probablity distribution. The marginalized parameter values, their uncertainties and the ranges over which we adopted a uniform prior are given in Table \ref{tbl:fit}.

We note that the formal uncertainties in $\Vgsr$ obtained from the H3 spectra are $<1 \,\kms$ for the \sgr{} members, with a median of $0.24\,\kms$.  Repeat observations have demonstrated that the quoted errors are under-estimated by a factor of two \citep{Conroy19a}.

The trends of mean velocity and velocity dispersion inferred from the H3 \sgr{} members with this model are shown as black lines and shaded regions in Figure \ref{fig:vel_fit}.  For comparison we show the trends inferred by \citet{Gibbons17}, who fit a combination of metal poor and metal rich components to SEGUE data (in their analysis ``metal-poor" referred to [Fe/H]$\approx-1.3$).  We did not find that residuals from the our inferred mean trend were corellated with metallicity. We also show the results of fitting our model to the most recently stripped LM10 simulation particles that fall within the H3 spatial window.  Finally, we indicate the velocity dispersions measured in the leading stream by \citet{Majewski04} and \citet{Monaco07}; the latter was used to constrain the LM10 model.  At its lowest point -- where projection effects are smallest -- the velocity dispersion in the trailing stream that we infer ($\sim 10 \,\kms$) is larger by $\sim 4-5\,\kms$ than we recover from the LM10 mock particles treated in the same way.

We infer diffuse fractions in raw number counts of $10\pm2$\% and $41\pm4$\% for $\Lambda_{\rm Sgr} < 150^\circ$ and $\Lambda_{\rm Sgr} > 200^\circ$ respectively. We use the cold stream models to identify members of the diffuse component as stars $>2\sigma$ from the mean $\Vgsr$ at any longitude. We have also computed marginalized posterior probabilities for membership in each component and used this to assign stars to cold and diffuse populations; the results are very similar.

\begin{figure*}
\includegraphics[width=\textwidth]{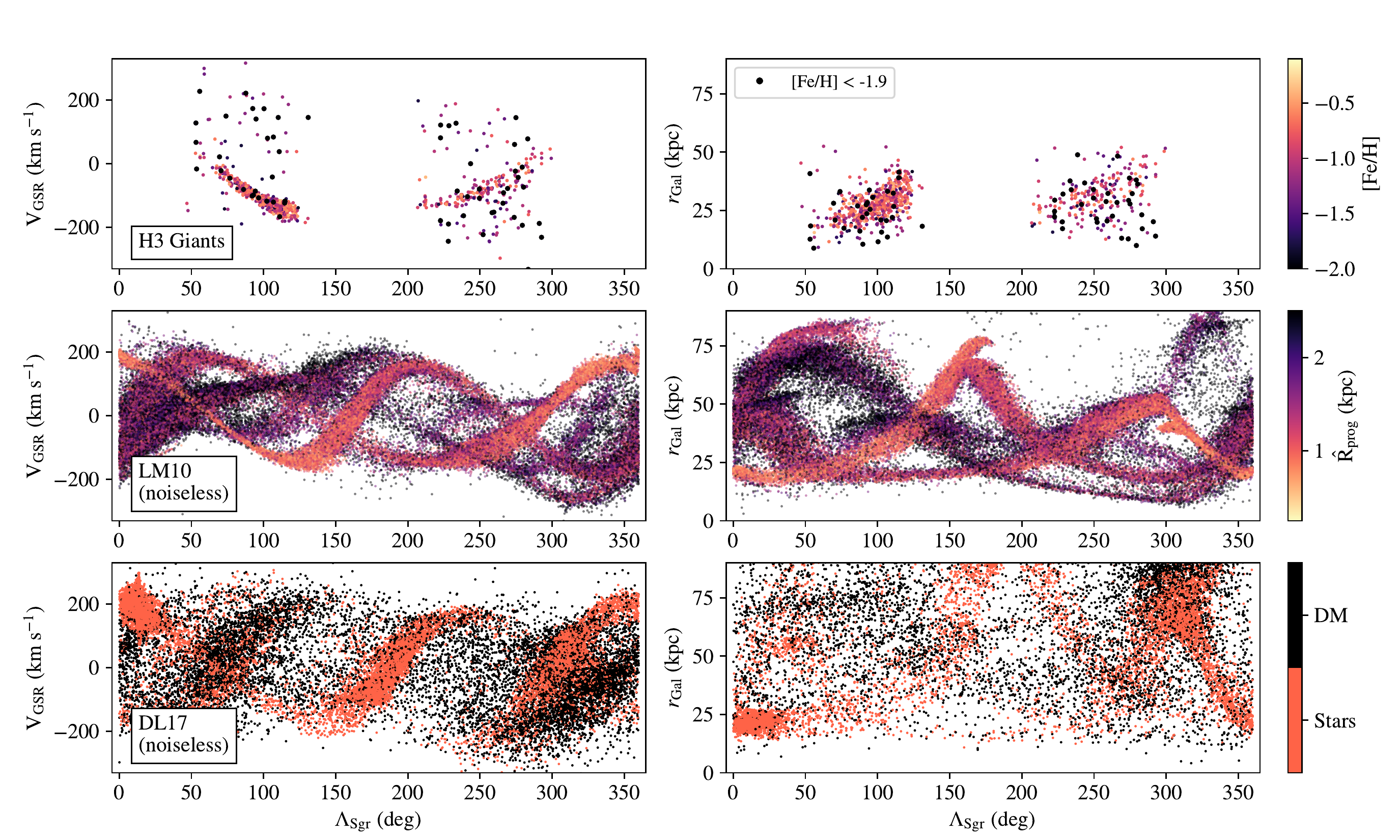}
\caption{
  $\Vgsr$ (left panels) and $r_{\rm Gal}$ (right panels) as a function of \sgr{} stream longitude.
  The top panels show the \sgr{} selected stars in H3 color-coded by metallicity.  The middle panels show a random subset of the \citet{LM10} model, color-coded by $\hat{\rm R}_{\rm prog}$, the approximate location within the original progenitor system. The bottom panels show the DL17 simulation, including both stars (orange points) and dark matter particles (black points).
\label{fig:full_mocks}}
\end{figure*}

\subsection{Stream Kinematics vs. Metallicity}

In this section we explore the kinematics of the \sgr{} stream as a function of metallicity.

Figure \ref{fig:ely} shows the orbital energy ($E_{\rm tot}$) as a function of the $y-$component of angular momentum ($\Ly$).  The left panel shows all H3 giants in grey and the \sgr{} members in black.  Members show up very clearly as a spur in the negative $\Ly$ direction \citep[see also][]{Hayes20}.  Moreover, two parallel diagonal sequences are clearly visible within the \sgr{} sample; these correspond to the leading and trailing arms \citep[see also][]{Li19}.  In the right panel we show only stars with [Fe/H]$<-1.9$.  This selection was chosen based on the break in the MDF in Figure \ref{fig:mdf}.  The spur is still clearly visible, which provides visual confirmation that the low metallicity population in Figure \ref{fig:mdf} is genuinely associated with the \sgr{} stream (see also Appendix \ref{apx:unc}).

In Figure \ref{fig:vgsrfeh} we show $\Vgsr$ as a function of the longitude along the stream, $\Lambda_{\rm Sgr}$.  The three panels correspond to three metallicity bins, with the most metal-rich at the top.  The bins were chosen to correspond to features in the MDF (see Figure \ref{fig:mdf}).  Stars are classified as belonging to a kinematically cold or diffuse component as described in Section \ref{s:vel_fit}.  The most metal-rich bin shown in the top panel reflects the ``conventional" view of the \sgr{} stream, e.g., as seen in M giant tracers \cite[e.g.,][]{Belokurov14}. Specifically, the metal-rich stream is kinematically cold, with $\sigma_v\lesssim 20\,\kms$ (see Section \ref{s:vel_fit}).

Remarkably, a kinematically diffuse component emerges at lower metallicities.  The diffuse component first appears at [Fe/H]$<-0.8$ (middle panel), and is the dominant component at [Fe/H]$<-1.9$ (bottom panel).  In fact, at the lowest metallicities the \sgr{} stream is barely identifiable in $\Vgsr$ space.  Recall that these stars are nonetheless very clearly members of the \sgr{} stream in angular momentum space (Figure \ref{fig:ely} and Figure \ref{fig:kin_unc} below).  The diffuse component comprises 24\% of the stars in the $-1.9<$[Fe/H]$<-0.8$ bin and 69\% of the stars at [Fe/H]$<-1.9$.

The distribution of metallicities for the cold and diffuse components is shown in Figure \ref{fig:mdfvel}.  Here the \sgr{} members are separated by their position along the stream ($\Lambda_{\rm Sgr}<140^\circ$ and $\Lambda_{\rm Sgr}>200^\circ$), and whether they are associated with the kinematically cold or diffuse components.  For the cold components, $\Lambda_{\rm Sgr}<140^\circ$ corresponds to the trailing arm, while $\Lambda_{\rm Sgr}>200^\circ$ corresponds to the leading arm.  Within the cold component the trailing arm is more metal-rich than the leading arm (median [Fe/H] of $-0.8$ compared to $-1.0$).   A more metal-rich trailing arm has been noticed before \citep[e.g.,][]{Carlin18, Hayes20}.  Models predict that at the range of stream longitudes sampled here the trailing arm has been more recently stripped than the leading arm, and so a more metal-rich trailing arm might arise if there was a steep metallicity gradient within the \sgr{} dwarf galaxy, as suggested by several authors \citep{Chou07, LM10, Hayes20}.

The MDFs of the kinematically diffuse components are shown in the bottom panels of Figure \ref{fig:mdfvel}.  Overall, the diffuse population has a much larger fraction of metal-poor stars than the kinematically cold population.  Given the small numbers of stars, it is difficult to discern any differences in the MDFs of the diffuse populations along the stream longitude.  However, there is tentative evidence that the low metallicity stars are not simply the tail of the distribution but appear as distinct components.  Additional data should clarify this issue.


\begin{figure*}[!t]
\includegraphics[width=\textwidth]{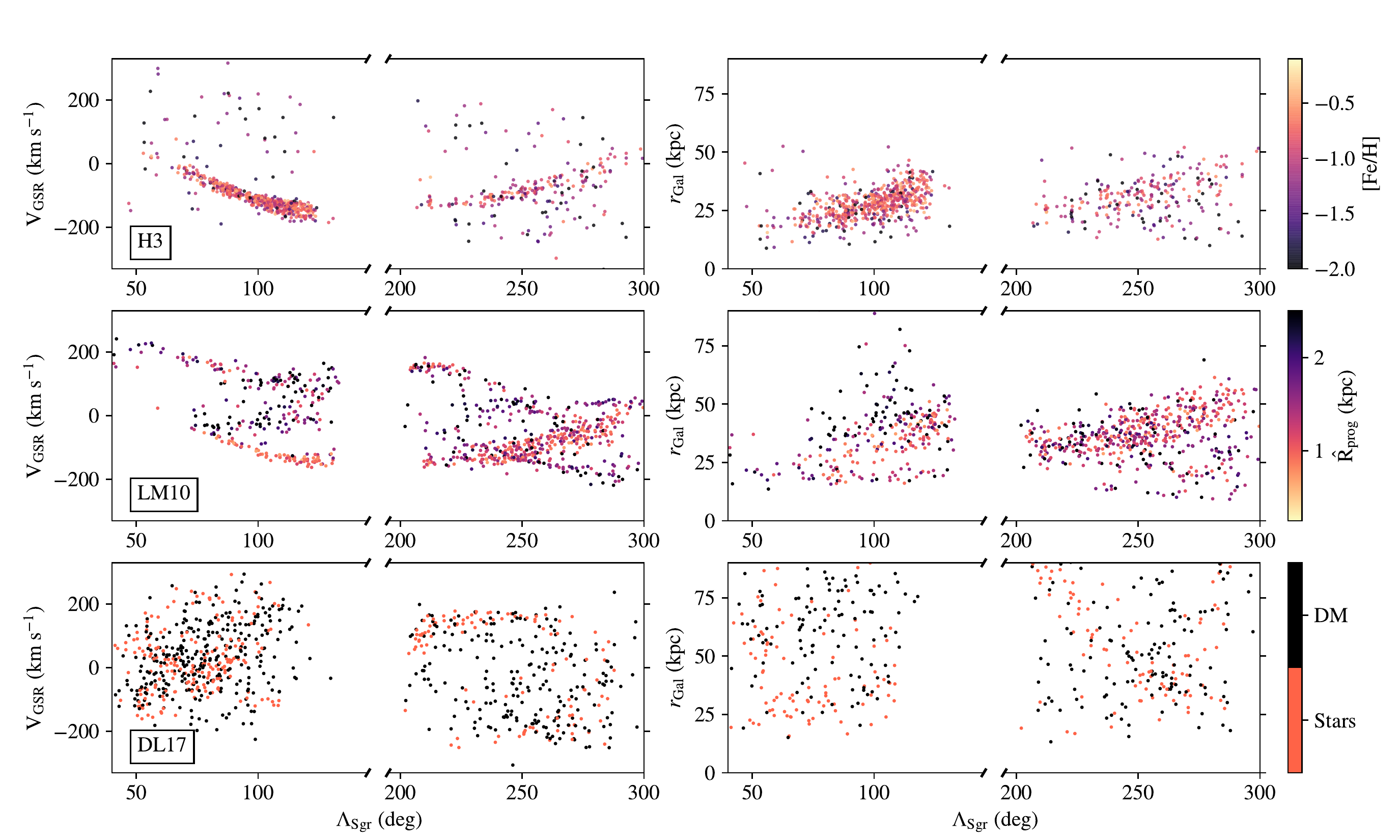}
\caption{As in Figure \ref{fig:full_mocks}, now with the H3 spatial selection window and error model applied to the LM10 and DL17 simulations.  For LM10 the H3 magnitude limit is also applied, while in the case of DL17, no attempt was made to simulate the magnitude limit in H3, and so the particles in the lower panel have a more extended distribution in $r_{\rm Gal}$ than in the other panels.
\label{fig:franken}}
\end{figure*}

\begin{figure*}[!t]
\includegraphics[width=\textwidth]{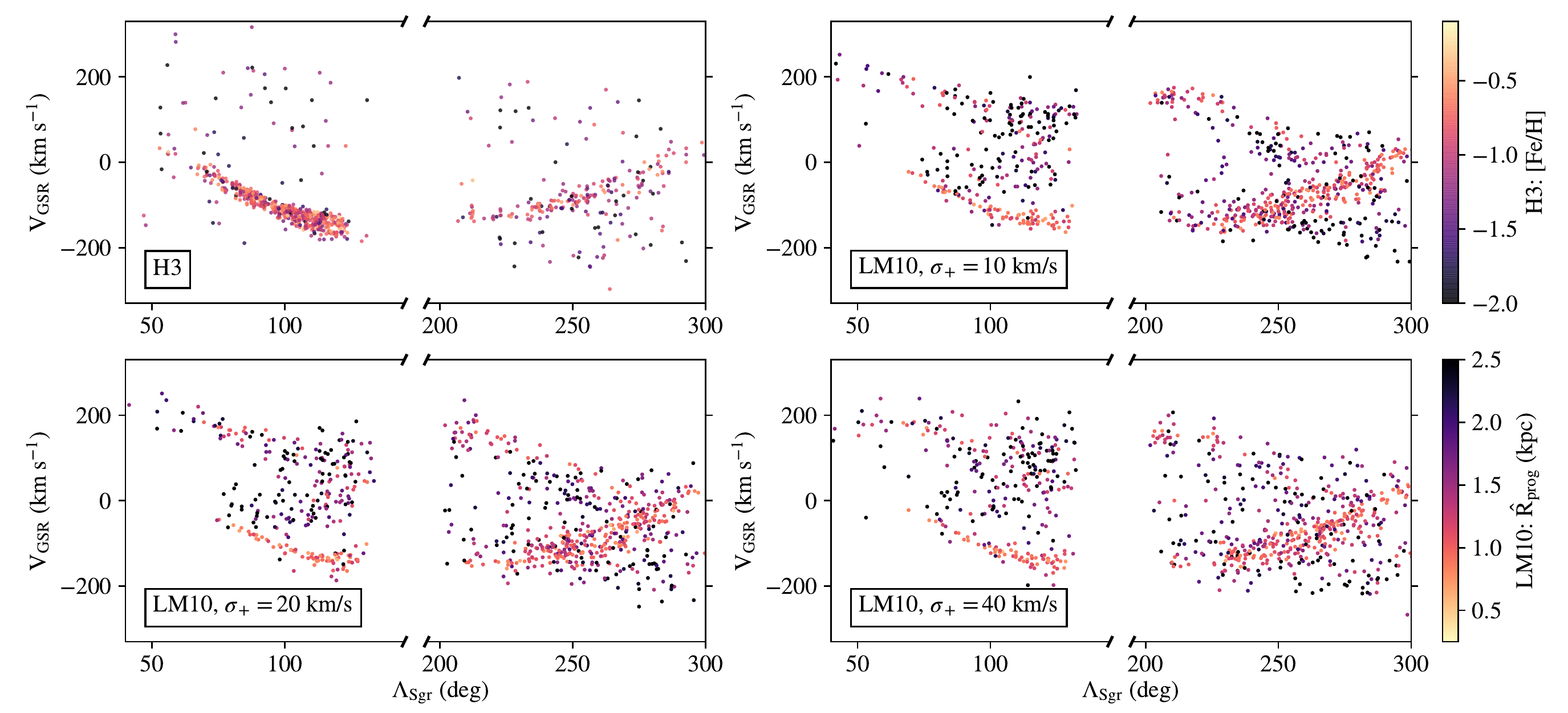}
\caption{ $\Vgsr$ as a function of \sgr{} stream longitude for \sgr{} selected stars in H3 colored by metallicity (top left panel) and for LM10 with the H3 selection function and error model applied and colored by $\hat{\rm R}_{\rm prog}$ (other panels).  In each of the LM10 panels an additional velocity dispersion is added to particles with $\hat{\rm R}_{\rm prog}>1.2$ kpc.  The default LM10 model produces debris at $\Vgsr>100\ \kms$ that is much colder than our data, while increasing the dispersion by $20$-$40\ \kms$ results in a somewhat better match to the data.
\label{fig:extra_sigma}}
\end{figure*}

\begin{figure*}
\includegraphics[width=\textwidth]{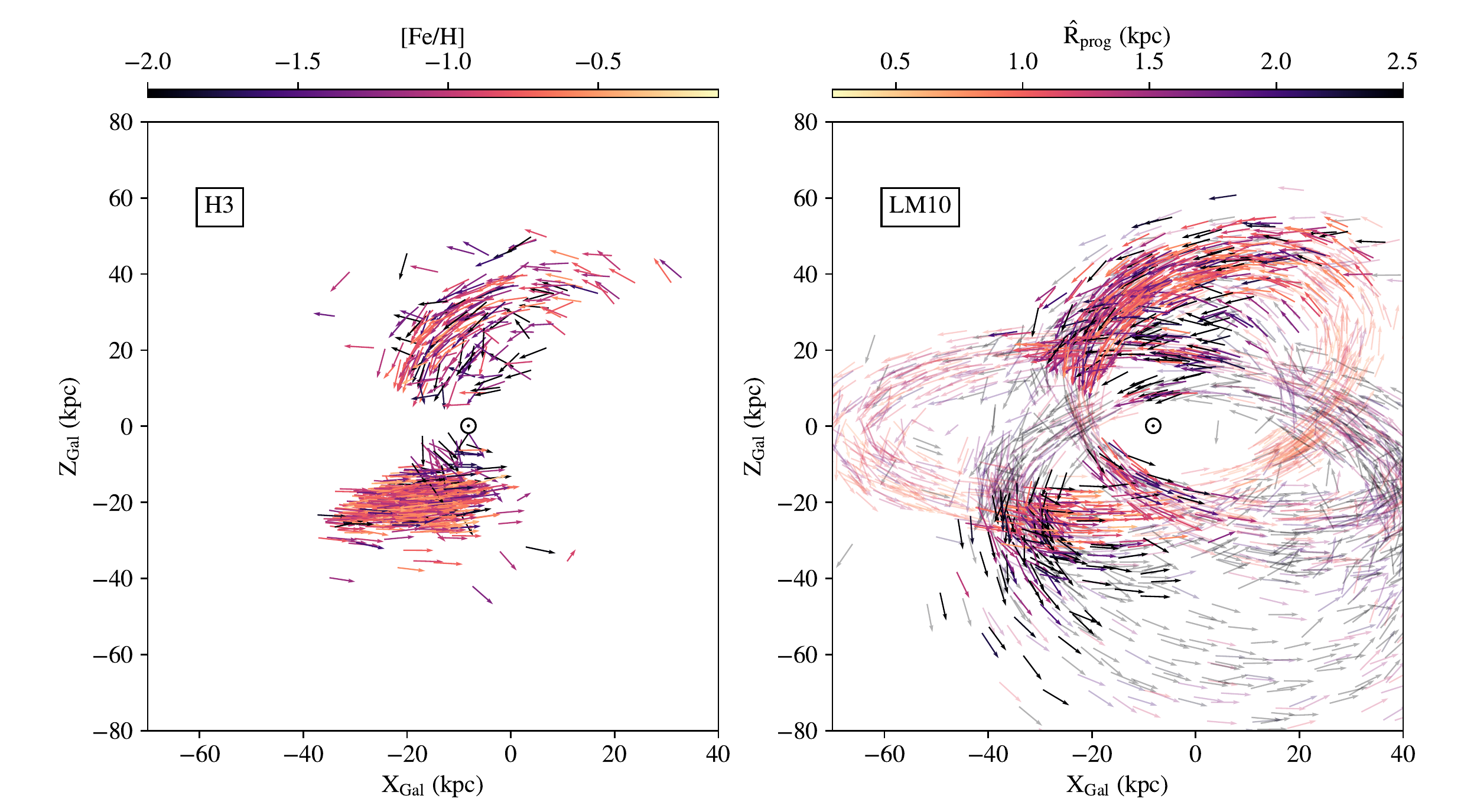}
\caption{Distribution of \sgr{} members in Galactocentric $X-Z$ coordinates.  Arrows indicate the direction of motion, and the length of the arrow is proportional to the velocity projected on the $X-Z$ plane.  In the left panel, H3 stars are color-coded by metallicity, while in the right panel LM10 stars are color-coded by $\hat{\rm R}_{\rm prog}$, the approximate location within the original progenitor system.  Furthermore, in the right panel, stars outside of the H3 footprint and magnitude range are shown as transparent arrows.  The position of the Sun is marked with a solar symbol.
\label{fig:quiver}}
\end{figure*}

\section{Comparison to Models}
\label{s:comp2models}

We now turn to a comparison between the data and the \sgr{} stream models of LM10 and DL17.

We begin by returning to Figure \ref{fig:lylz}, in which the models and data are shown in angular momentum space.  We note first that the stream stars in H3 are offset with respect to the observed remnant \citep{Fritz18}.  This behavior is not observed in LM10 because angular momentum transfer was not allowed between the low mass \sgr{} progenitor and the rigid Milky Way halo. There is an unobserved spur of debris in LM10 at positive $\Lz$ and more negative values of $\Ly$ compared to the bulk of their model that is associated with stars that became unbound early in the interaction, the most uncertain regime of that model. In the DL17 model, in which the dynamic host halo enabled angular momentum transfer via dynamical friction, two modes offset from the modeled remnant are visible, corresponding to the leading and trailing streams.

Figure \ref{fig:full_mocks} compares the H3 velocities and Galactocentric distances to the LM10 and DL17 models as a function of stream longitude ($\Lambda_{\rm Sgr}$).  The H3 stars are color-coded by metallicity, and the most metal-poor stars are highlighted as larger black symbols.  The LM10 model points are color-coded by their mean orbital radius within the progenitor system, which very roughly correlates with the time at which a particle became unbound from \sgr{}.  For the DL17 model we show the locations of both the stripped stars (orange) and the dark matter particles (black).  The purpose of showing both the stars and dark matter is to compare the stream morphology of the colder stellar component to the more diffuse dark matter component.  For both the LM10 and DL17 models, we display the full, noiseless simulation particle data.

The LM10 models were tuned in part to reproduce the observed cold component, so it is not surprising that those models reproduce that aspect of the data.  The DL17 models were not tuned to the same degree, and so there is somewhat less agreement with the observed cold component.

It is intriguing that both models predict populations at the same approximate stream longitude coordinates as the well-studied cold components (e.g., at $50^\circ<\Lambda_{\rm Sgr}<130^\circ$ and $200^\circ<\Lambda_{\rm Sgr}<300^\circ$) but offset in $\Vgsr$.  In LM10 these structures at different $\Vgsr$ were stripped at earlier times from the outer regions of the progenitor system, while in DL17 there is overlapping debris both from the stellar and dark matter components.  From these two models we can infer that material stripped at earlier times will in general not lie in the same regions of $\Vgsr-\Lambda_{\rm Sgr}$ and $r_{\rm Gal}-\Lambda_{\rm Sgr}$ as the more recently stripped material, in spite of the fact that all of this debris occupies a similar region in angular momentum space (see Figure \ref{fig:lylz}).

A more direct comparison between the data and models is provided in Figure \ref{fig:franken}, where we have attempted to create an H3-like survey from the LM10 and DL17 simulation data, as described in Section \ref{sec:simulations}.  In this figure the simulated data are down-sampled to produce the same number of points as in the H3 panel.  It is noteworthy that the cold component at $50^\circ<\Lambda_{\rm Sgr}<130^\circ$ in DL17 is greatly diminished (compare with Figure \ref{fig:full_mocks}).  This is due to the fact that the DL17 model does not project into the correct on-sky position of the \sgr{} stream.  Furthermore, the trailing arm in LM10 is too cold compared to the data (see also Figure \ref{fig:vel_fit}).  This suggests that the mass for \sgr{} adopted in LM10 is too low \citep[see also][]{Gibbons17}.

Turning to the diffuse component in H3, there is some general correspondence with the LM10 and DL17 models, in the sense that both models predict additional debris at $50^\circ<\Lambda_{\rm Sgr}<130^\circ$, $\Vgsr>0\,\kms$, and $200^\circ<\Lambda_{\rm Sgr}<300^\circ$ at both $\Vgsr\approx-200\,\kms$ and $\Vgsr>0\,\kms$ \citep[see also][]{Yang19}.  However, the LM10 predictions appear in general much colder than the observations.  We explore this further in Figure \ref{fig:extra_sigma} where we have convolved the LM10 velocities by an additional dispersion of $\sigma_+=10,20,40\ \kms$ for stars with $\hat{\rm R}_{\rm prog}>1.2$ kpc.  Visual comparison between the data and these artificially-broadened LM10 models suggests that the data at $\Vgsr>0\,\kms$ are $20-40\ \kms$ more diffuse than the default LM10 models.  This is in contrast to the cold wraps, in which the LM10 model is only $\approx5\kms$ colder than the data (see Figure \ref{fig:vel_fit}).  Returning to Figure \ref{fig:franken}, the DL17 model predicts a high degree of diffuse structure, if we associate some stars with the dark matter distribution (black points).

We also note that the relative density of points in the cold and diffuse components is quite different between the data and the LM10 model (recall that LM10 has been downsampled to the same total number of H3 \sgr{} members).  In particular, at $50^\circ<\Lambda_{\rm Sgr}<130^\circ$ there are far more stars in the cold component in the data compared to the LM10 model.  This could mean that the radial density profile in the LM10 progenitor model is too shallow.

In general, the conclusion from this comparison is that neither model accurately predicts both the locations and large spread in velocity of the metal-poor component.  However, by combining insights from both models, we suggest that the diffuse metal-poor component observed in the data is associated with older wraps of the \sgr{} stream that are probing the outer regions of the progenitor system.

Finally, in Figure \ref{fig:quiver} we compare \sgr{} members in H3 to the LM10 model in configuration space.  Because the orbit of \sgr{} is closely aligned in the $Y$-plane, we show stars in the $X-Z$ plane.  Arrows show the direction of motion, with the length normalized by the magnitude of the velocity.   H3 stars are color-coded by metallicity, while LM10 points are color-coded by their mean orbital radius within the progenitor.  For LM10, points that would lie within the H3 Survey footprint and magnitude limit are shown as solid, and the rest are shown as transparent.

There are several interesting features in Figure \ref{fig:quiver}.  First, the continuation of the leading stream at ($X_{\rm Gal}$, $Z_{\rm Gal}$) $=$ ($-10,-20$) kpc that is heading towards negative $Z_{\rm Gal}$ corresponds approximately to the leading arm in LM10 at the same coordinates.  However, in the LM10 model, this portion of the leading arm has a larger velocity component in the $X_{\rm Gal}$ direction, and appears much colder.  The earlier portion of the leading arm extends to greater $Z_{\rm Gal}$ than seen in H3 (50 vs. 40 kpc).  The stars at ($-20,20$) kpc moving toward negative $X_{\rm Gal}$ are likely a continuation of the trailing arm wrapping back around the Galaxy, as previously noted by \citet{Yang19}.  Finally, we do not detect the older predicted wrap in LM10 at ($-30,-30$) kpc with a significant $-Z_{\rm Gal}$ velocity component.

In spite of the known shortcomings of the LM10 model, it is still widely used owing to its ability to match in detail many of the features of the cold debris at $<50$ kpc.  There would be significant value in an updated version of an LM10-style model that is able to more accurately reproduce the extended debris at larger radius.  In light of the results presented here, there would also be value in considering multi-component models of the progenitor, for example a compact main body and an extended metal-poor stellar halo.

\section{Discussion}
\label{sec:discussion}

In this paper we have combined {\it Gaia} and H3 data to identify $\Nsgr$ \sgr{} members based on a simple selection in angular momentum space.  Owing to the H3 survey design, the resulting sample is nearly unbiased with respect to metallicity.  This selection allowed us to identify a population of metal-poor stars ([Fe/H]$<-1.9$) associated with the \sgr{} stream that is both offset and more diffuse in kinematic space compared to the metal-rich component.  By comparing to simulations of the \sgr{} stream, we infer that this metal-poor component was likely stripped from the \sgr{} progenitor at earlier times than the more metal-rich colder component.

These results support a picture in which this metal-poor component of \sgr{} represents a population of stars within the progenitor system at larger radius and perhaps with higher velocity dispersion, compared to the main body.  Such a population could be considered the stellar halo of the \sgr{} dSph galaxy.

Extended structures with distinct stellar populations are common in star-forming dwarf galaxies in the Local Volume \citep[see][and references therein]{Stinson09}. Many nearby dwarf galaxies show some evidence for stellar halo-like populations, including Sculptor \citep{Tolstoy04}, Fornax \citep{Battaglia06}, Sextans \citep{Battaglia11}, and Ursa Minor \citep{Pace20}.  These galaxies have clear metallicity gradients with a kinematically hotter, more metal-poor population extending to larger radius than the colder, more metal-rich population.  M33, the largest satellite in the Local Group, also shows clear evidence of a metal-poor population at large radius \citep{Cioni09}.  The Small and Large Magellanic Clouds (SMC and LMC) have relatively shallow metallicity gradients \citep{Cioni09}.  However, RR Lyrae in the LMC do suggest the presence of a kinematically-hot, metal-poor stellar halo \citep{Borissova06}.  In deep optical imaging \citet{Kado-Fong20} find round stellar outskirts, suggestive of stellar halos, to be ubiquitous in $M_\ast \sim 10^{9} M_\odot$ galaxies. Focusing on Sagittarius, it has long been recognized that the metallicity gradient along the {\it cold} leading and trailing streams, and the metallicity difference between the streams and the remnant, implies a very steep metallicity gradient within the progenitor system \citep[e.g.,][]{Bellazzini06, Chou07, LM10, Hayes20}.

The origin of dwarf stellar halos is unclear.  Dwarf mergers are predicted to be common in the early hierarchical growth of structure predicted by cold dark matter cosmology \citep[e.g.,][]{Deason14}, though the decreasing stellar mass to halo mass ratios at lower hower halo masses and realistic hydrodynamical simulations predict that such mergers account for a small fraction of the stellar mass in present-day dwarfs \citep[e.g.,][]{Purcell07, Fitts18}.  A shell structure in deep imaging of the Fornax dSph has been interpreted as evidence of recent accretion of a smaller dwarf system \citep{Coleman04}.  Using simulations, \citet{BenitezLlamblay16} and \citet{Genina19} argue for several pathways to produce the metal-rich/metal-poor dichotomy in dwarfs, all of which are related to a history of mergers within the system \citep[see also][]{Revaz18}.  \citet{Kawata06} attempt to explain the metallicity dichotomy in Sculptor solely by dissipative collapse at high redshift.  While this model is able to produce a metallicity gradient, it does not produce a substantially hotter metal-poor component, in contrast with the data.  The formation of extended, older, stellar halos in dwarfs was found by \citep{Stinson09} to be possible via {\it in situ} processes including disk sloshing and outflows, though the kinematic signatures were not explored. Finally, \citet{El-Badry16} argue that rapid potential fluctuations induced by stellar feedback can efficiently re-distribute stellar populations resulting in (modestly) negative metallicity gradients.  This last scenario could potentially also produce a kinematically hotter population at large radius, but it is unclear if it can generate the steep observed metallicity gradients.

The kinematic offset between the metal-poor and metal-rich populations, combined with insights from simulations, is our strongest argument in favor of the metal-poor stars belonging to a halo-like population.  The diffuse kinematics support this picture, but there are alternative explanations for the diffuse appearance in kinematic space (Figure \ref{fig:extra_sigma}).  One possibility is that we are seeing multiple older cold wraps that are overlapping and simply appear diffuse.  This seems unlikely based on comparison to LM10 (see Figure \ref{fig:franken}), and we stress that observational uncertainties in $\Vgsr$ are solely a function of the measured radial velocity and hence are very small ($\approx1\ \kms$).  Another possibility is that these older wraps were cold when stripped from the progenitor system and  subsequently dynamically heated.  Given the long orbital times in the outer halo this too seems unlikely, but detailed simulations are required to clarify this option.  Finally, the kinematically-offset stars may be due to one or more satellites of Sagittarius that were subsequently tidally destroyed in the Galactic potential.

We now place our results in the broader context of the \sgr{} system.  We measured a mean metallicity of the entire stream of $\langle$[Fe/H]$\rangle_{\rm stream}=-0.99$.  Using APOGEE data, \citet{Hayes20} measured the metallicity of the remnant of $\langle$[Fe/H]$\rangle_{\rm rem}=-0.57$.  \citet{Niederste-Ostholt10} has estimated the total luminosity of the \sgr{} system (stream plus remnant) to be $L\approx1.1\times10^8\, L_{\odot}$, and they estimated that 70\% of the luminosity is in the stream.  We therefore take a weighted average of the stream and remnant metallicities to arrive at an average \sgr{} system metallicity of $\langle$[Fe/H]$\rangle_{\rm system}=-0.86$.  Assuming $M/L_V=5$ would imply a stellar mass of $M_\ast=5.5\times10^8\, M_\odot$.  The mass-metallicity relation for Local Group dwarfs determined by \citet{Kirby13} predicts a metallicity of $-0.88$ for this mass, in remarkable agreement with our estimated remnant metallicity.  The inferred stellar mass of Sagittarius lies in between the stellar masses of the SMC \citep[$3\times10^8\, \Msun$;][]{Stanimirovic04} and LMC \citep[$3\times10^9\, \Msun$;][]{vanderMarel02}.

\begin{figure*}[t!]
\includegraphics[width=\textwidth]{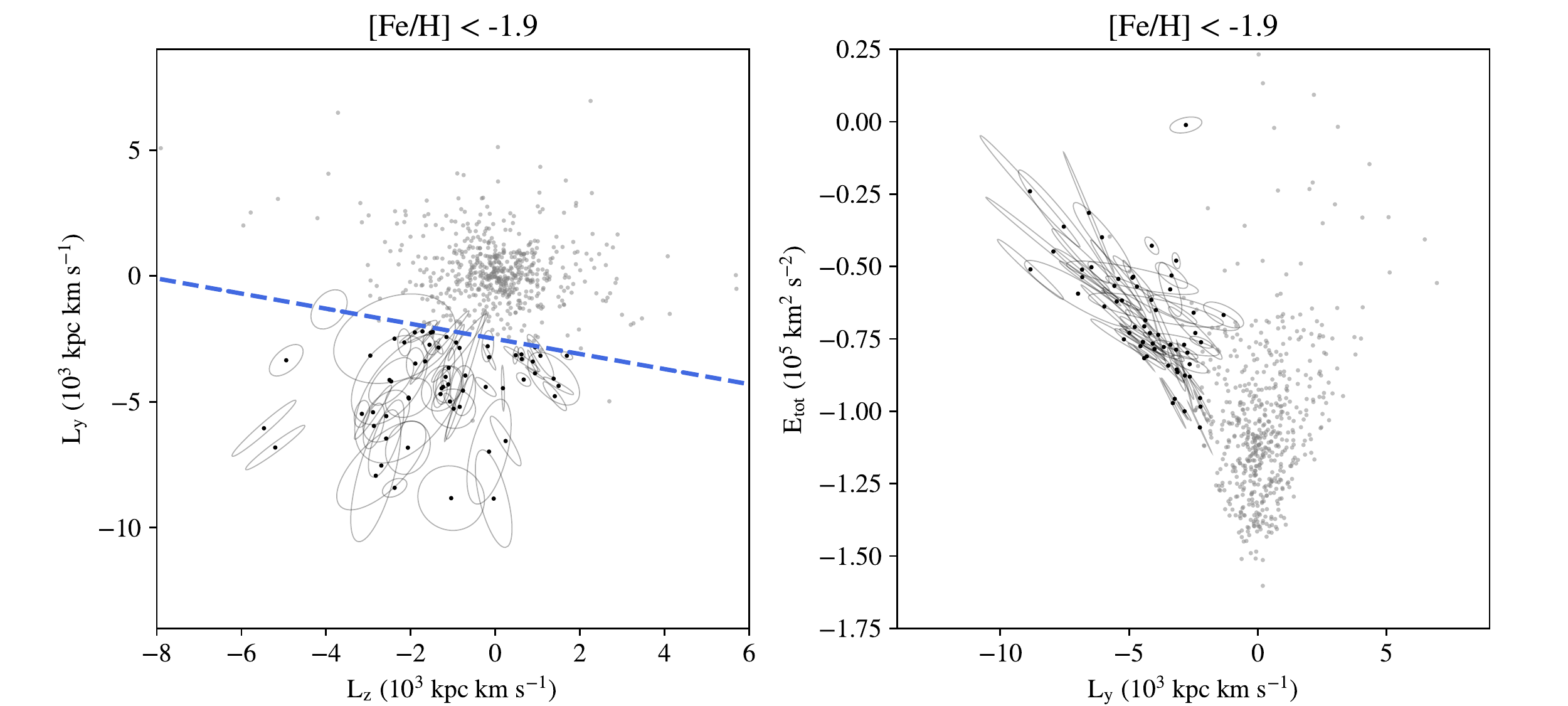}
\caption{Distribution of low metallicity H3 giants in both the $\Lz-\Ly$ plane (left panel) and the $E-\Ly$ plane (right panel).  The blue dashed line in the left panel shows the \sgr{} selection criteria.  \sgr{} members with [Fe/H]$<-1.9$ are shown as darker symbols.  Ellipses show the highly correlated uncertainties of the low-metallicity \sgr{} members in both of these planes.
\label{fig:kin_unc}}
\end{figure*}

The total mass of \sgr{} before infall is quite uncertain, with estimates ranging from $10^{9}-10^{11}\, M_\odot$ \citep[e.g.,][]{Jiang00,Helmi01, LM10, Lokas10, Purcell11, Gibbons17,Laporte18}.  Part of the challenge in constraining the mass lies in the fact that a wide range of initial masses can produce a comparable present-day remnant mass and location \citep[e.g.,][]{Gibbons17}.  Abundance matching halos to galaxies in a cosmological setting predicts a total mass of $M_{\rm halo}\approx10^{11}\, M_\odot$ \citep{Behroozi19} for our adopted stellar mass.  Another approach is to use the globular cluster (GC) system to estimate the total halo mass, as many authors have noted a strong power-law relation between the two \citep[e.g.,][]{Hudson14, Harris17}.  We have identified seven GCs as confidently associated with \sgr{} (see Appendix \ref{apx:globulars}).  Using the \citet{Harris96} catalog (2010 edition), and $M/L_V=2$, we estimate a total GC mass in the \sgr{} system of $M_{\rm GC}=3\times10^6\, M_\odot$.  Adopting the GC mass-to-halo mass ratio of $M_{\rm GC}/M_{\rm halo}=4\times10^{-5}$ from \citet{Hudson14} leads to an estimate of the \sgr{} progenitor halo mass of $M_{\rm halo}\approx 6\times 10^{10}\, M_\odot$.  A relatively ``heavy" \sgr{} progenitor mass would have implications for the predicted velocity dispersion of any associated stellar halo.

An important limitation to this work is the incomplete view of \sgr{} provided by the H3 Survey in terms of on-sky coverage.  The current H3 footprint is inhomogeneous \citep[see][]{Conroy19a}, so we are likely missing important features of the \sgr{} system.  H3 will eventually homogeneously (and sparsely) cover the entire sky at $|b|>30^\circ$ and Dec.$>-20^\circ$.  However, even the final dataset cannot provide a complete all-sky view of \sgr{}.  Combining spectroscopic surveys with large-area photometric surveys \citep[e.g.,][]{Sesar17, Antoja20}, will therefore continue to be essential to develop a complete view of the \sgr{} system.

An additional limitation to the present study is the impact of distance and proper motion uncertainties on the selection of \sgr{} members.  Both of these quantities are uncertain at the $\approx10$\% level.  While the spectrophotometric distance uncertainties are unlikely to substantially improve, the proper motions are expected to become much more precise in future {\it Gaia} data releases.  The current uncertainties could result in some degree of contamination in the membership selection, although we would expect any contamination to occur independent of metallicity (see further discussion in Appendix \ref{apx:unc}).  Even with perfect measurements there will likely be some contamination from the background Milky Way populations. Bringing other information to bear on the selection, such as chemistry, could be useful in this case.

The metal-poor stars belonging to \sgr{} comprise 11\% of the total sample of metal-poor giants in the H3 Survey.  Unlike the metal-rich \sgr{} population, these stars do not appear as cold structures either on-sky or in velocity-position space.  The identification of this population is a testament to the power of considering conserved quantities such as angular momenta when identifying debris in the Galactic halo.

\acknowledgments

YST is supported by the NASA Hubble Fellowship grant HST-HF2-51425.001 awarded by the Space Telescope Science Institute.  CC acknowledges support from the Packard Foundation.

We thank the Hectochelle operators Chun Ly, ShiAnne Kattner, Perry Berlind, and Mike Calkins, and the CfA and U. Arizona TACs for their continued support of the H3 Survey. Observations reported here were obtained at the MMT Observatory, a joint facility of the Smithsonian Institution and the University of Arizona. Computations for this study were run on the FASRC Cannon cluster supported by the FAS Division of Science Research Computing Group at Harvard University.

\software{Minesweeper \citep{Cargile19},
          dynesty \citep{dynesty},
          gala \citep[v1.1][]{gala1, gala2, bovy15},
          Astropy \citep[v4.0][]{astropy1, astropy2},
          NumPy \citep{numpy},
          SciPy \citep{scipy},
          matplotlib \citep{matplotlib},
          IPython \citep{ipython}
         }

\begin{appendix}

\begin{figure*}[t!]
\includegraphics[width=\textwidth]{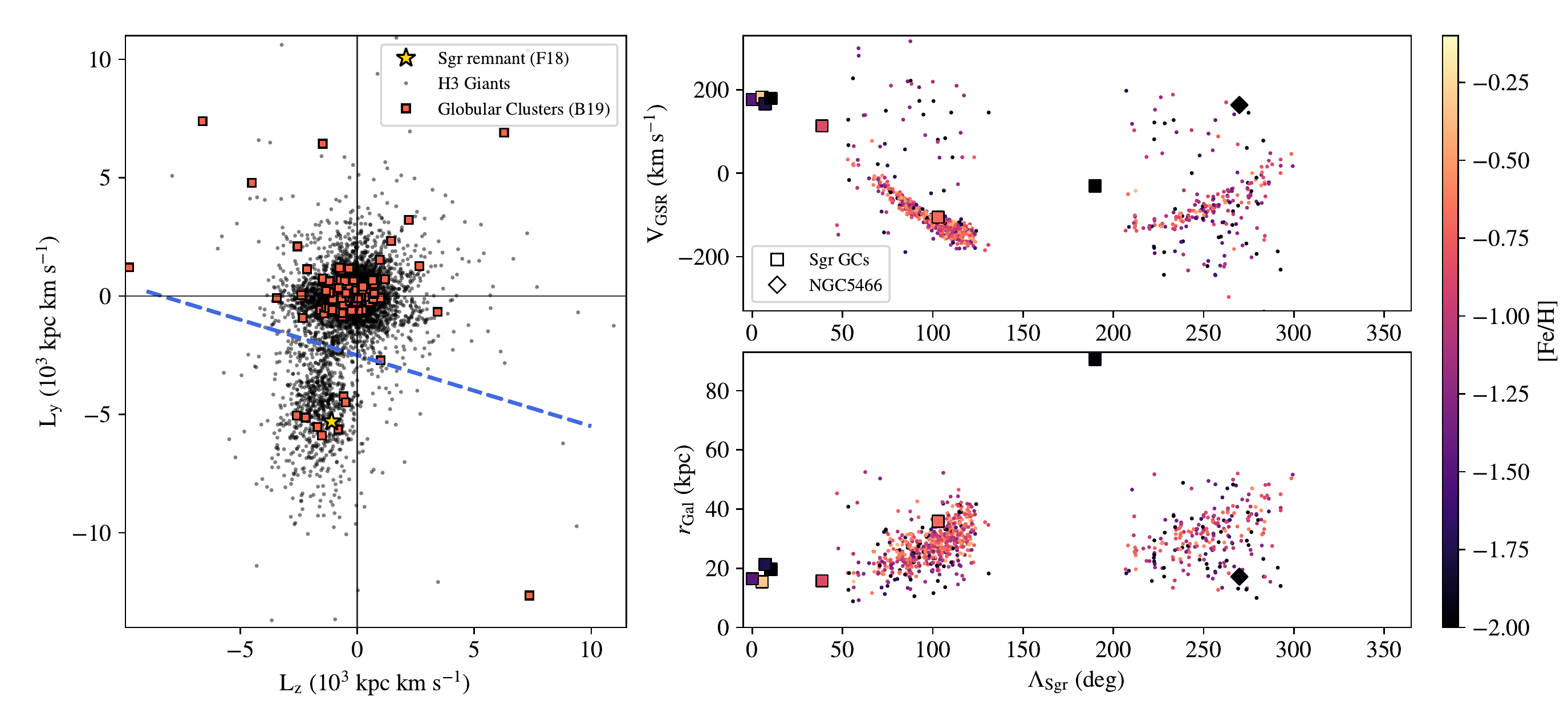}
\caption{Distribution of all GCs from \citet{Baumgardt19} in phase space.  Left panel: distribution in $\Lz-\Ly$, comparing the locations of the GCs to the H3 sample.  The seven \sgr{} GCs (NGC 2419, NGC 6715, Pal 12, Terzan 7, Terzan 8, Arp 2, and Whiting 1) are clearly clustered around the \sgr{} remnant \citep[star symbol;][]{Fritz18} in angular momentum space.  NGC 5466 lies on the selection boundary.  Right panels: Comparison of \sgr{} GCs to H3 \sgr{} members in $\Vgsr$ and Galactocentric distance, as a function of stream longitude.  Symbols are color-coded by metallicity.
\label{fig:globulars}}
\end{figure*}

\section{Uncertainties in Phase Space Quantities}
\label{apx:unc}

It is important to consider the uncertainties on the kinematic quantities that we are using to select \sgr{} members.  These uncertainties are highly correlated. We have propagated uncertainties by computing the relevant kinematic quantities from a number of fair samples of the posterior distributions for the phase-space coordinates of each star.  For heliocentric distances and radial velocities we take samples from the posteriors computed with \minesweeper{}. For proper motions we sample from Gaussians described by the values and uncertainties provided by the \emph{Gaia} DR2.  The uncertainties in celestial coordinates are negligible.  For the \sgr{} members, the error budget is dominated by two terms: the $\approx10$\% distance uncertainty, and the approximately comparable uncertainty in the proper motions.

We then estimate the covariance matrix of $\Ly$, $\Lz$ (and $E_{\rm tot}$) for each star from these posterior samples.  While displaying the uncertainty ellipses for all stars is challenging, in Figure \ref{fig:kin_unc} we show the uncertainty ellipses for an important subset of H3 stars:  \sgr{} members with $[{\rm Fe/H}] < -1.9$.  This figure demonstrates that the uncertainties can in some cases be significant, but they do not compromise the identification of \sgr{} members at low metallicity.

\section{Sagittarius Globular Clusters}
\label{apx:globulars}

The origin of the Galactic GC population has been the subject of much debate.  Recently it has become clear that many, if not most of the GCs are associated with accreted galaxies in the halo \citep[e.g.,][]{Bellazzini03, Law10b, Myeong19, Massari19, Kruijssen20}.  In this Appendix we revisit the question of which GCs are associated with \sgr{} in light of our $\Lz-\Ly$ selection (Figure \ref{fig:lylz}).

We use the catalog of GC proper motions, distances, and velocities from \citet{Baumgardt19} and compute associated angular momenta and projections into the \sgr{} orbital plane.  We show the distribution of all 154 GCs in $\Lz-\Ly$ in Figure \ref{fig:globulars}.  There are seven GCs clearly associated with \sgr{} in angular momentum space: NGC 2419, NGC 6715, Pal 12, Terzan 7, Terzan 8, Arp 2, and Whiting 1.  Four of these have long been associated with the core of the \sgr{} dSph (NGC 6715 [M54], Terzan 7, Terzan 8, and Arp 2).  Whiting 1, Pal 12, and NGC 2419 have also previously been suggested to be associated with \sgr{} \citep[e.g.,][]{Newberg03, Law10b, Belokurov14, Massari19, Bellazzini20}.  A number of other clusters have been proposed as being associated with \sgr{} that do not meet our selection criterion: Berkeley 29, NGC 5634 and NGC 5053 \citep{Law10b}, NGC 5824 \citep{Massari19}, NGC 5634 and NGC 4147 \citep{Bellazzini20}.  Finally, there is one cluster, NGC 5466, that falls right on top of the selection boundary, and to our knowledge has not previously been associated with \sgr{}.

The right panels of Figure \ref{fig:globulars} show the seven strong candidates and NGC 5466 in the $\Vgsr-\Lambda_{\rm Sgr}$ and $r_{\rm Gal}-\Lambda_{\rm Sgr}$ planes.  The points are color-coded by metallicity, and are plotted along with the H3 \sgr{} members.  Six of the clusters are at $0^\circ < \Lambda_{\rm SGR}<100^\circ$ and are clearly associated with either the main body or the cold component.  NGC 2419, at $\Lambda_{\rm Sgr}\approx 190^\circ$, is associated with the apocenter of the cold trailing arm \citep{Newberg03, Belokurov14}.  Finally, NGC 5466, whose association with \sgr{} we hold as tentative, lies at $\Lambda_{\rm Sgr}\approx 270^\circ$ and is coincident with the diffuse metal-poor population.  We regard this association as suggestive and worthy of further investigation.

\end{appendix}


\end{CJK*}

\begin{thebibliography}{}
\expandafter\ifx\csname natexlab\endcsname\relax\def\natexlab#1{#1}\fi
\providecommand{\url}[1]{\href{#1}{#1}}

\bibitem[{{Alard}(1996)}]{Alard96}
{Alard}, C. 1996, \apjl, 458, L17

\bibitem[{{Alcock} {et~al.}(1997){Alcock}, {Allsman}, {Alves}, {Axelrod},
    {Becker}, {Bennett}, {Cook}, {Freeman}, {Griest}, {Guern}, {Lehner},
    {Marshall}, {Minniti}, {Peterson}, {Pratt}, {Quinn}, {Rodgers}, {Stubbs},
    {Sutherland}, \& {Welch}}]{Alcock97}
{Alcock}, C., {Allsman}, R.~A., {Alves}, D.~R., {et~al.} 1997, \apj, 474, 217

\bibitem[{{Antoja} {et~al.}(2020){Antoja}, {Ramos}, {Mateu}, {Helmi}, {Anders},
    {Jordi}, \& {Carballo-Bello}}]{Antoja20}
{Antoja}, T., {Ramos}, P., {Mateu}, C., {et~al.} 2020, \aap, 635, L3

\bibitem[{{Astropy Collaboration} {et~al.}(2013){Astropy Collaboration},
    {Robitaille}, {Tollerud}, {Greenfield}, {Droettboom}, {Bray}, {Aldcroft},
    {Davis}, {Ginsburg}, {Price-Whelan}, {Kerzendorf}, {Conley}, {Crighton},
    {Barbary}, {Muna}, {Ferguson}, {Grollier}, {Parikh}, {Nair}, {Unther},
    {Deil}, {Woillez}, {Conseil}, {Kramer}, {Turner}, {Singer}, {Fox}, {Weaver},
    {Zabalza}, {Edwards}, {Azalee Bostroem}, {Burke}, {Casey}, {Crawford},
    {Dencheva}, {Ely}, {Jenness}, {Labrie}, {Lim}, {Pierfederici}, {Pontzen},
    {Ptak}, {Refsdal}, {Servillat}, \& {Streicher}}]{astropy1}
{Astropy Collaboration}, {Robitaille}, T.~P., {Tollerud}, E.~J., {et~al.} 2013,
    \aap, 558, A33

\bibitem[{{Astropy Collaboration} {et~al.}(2018){Astropy Collaboration},
    {Price-Whelan}, {Sip{\H o}cz}, {G{\"u}nther}, {Lim}, {Crawford}, {Conseil},
    {Shupe}, {Craig}, {Dencheva}, {Ginsburg}, {VanderPlas}, {Bradley},
    {P{\'e}rez-Su{\'a}rez}, {de Val-Borro}, {Aldcroft}, {Cruz}, {Robitaille},
    {Tollerud}, {Ardelean}, {Babej}, {Bach}, {Bachetti}, {Bakanov}, {Bamford},
    {Barentsen}, {Barmby}, {Baumbach}, {Berry}, {Biscani}, {Boquien}, {Bostroem},
    {Bouma}, {Brammer}, {Bray}, {Breytenbach}, {Buddelmeijer}, {Burke},
    {Calderone}, {Cano Rodr{\'{\i}}guez}, {Cara}, {Cardoso}, {Cheedella},
    {Copin}, {Corrales}, {Crichton}, {D'Avella}, {Deil}, {Depagne}, {Dietrich},
    {Donath}, {Droettboom}, {Earl}, {Erben}, {Fabbro}, {Ferreira}, {Finethy},
    {Fox}, {Garrison}, {Gibbons}, {Goldstein}, {Gommers}, {Greco}, {Greenfield},
    {Groener}, {Grollier}, {Hagen}, {Hirst}, {Homeier}, {Horton}, {Hosseinzadeh},
    {Hu}, {Hunkeler}, {Ivezi{\'c}}, {Jain}, {Jenness}, {Kanarek}, {Kendrew},
    {Kern}, {Kerzendorf}, {Khvalko}, {King}, {Kirkby}, {Kulkarni}, {Kumar},
    {Lee}, {Lenz}, {Littlefair}, {Ma}, {Macleod}, {Mastropietro}, {McCully},
    {Montagnac}, {Morris}, {Mueller}, {Mumford}, {Muna}, {Murphy}, {Nelson},
    {Nguyen}, {Ninan}, {N{\"o}the}, {Ogaz}, {Oh}, {Parejko}, {Parley}, {Pascual},
    {Patil}, {Patil}, {Plunkett}, {Prochaska}, {Rastogi}, {Reddy Janga},
    {Sabater}, {Sakurikar}, {Seifert}, {Sherbert}, {Sherwood-Taylor}, {Shih},
    {Sick}, {Silbiger}, {Singanamalla}, {Singer}, {Sladen}, {Sooley},
    {Sornarajah}, {Streicher}, {Teuben}, {Thomas}, {Tremblay}, {Turner},
    {Terr{\'o}n}, {van Kerkwijk}, {de la Vega}, {Watkins}, {Weaver}, {Whitmore},
    {Woillez}, {Zabalza}, \& {Astropy Contributors}}]{astropy2}
{Astropy Collaboration}, {Price-Whelan}, A.~M., {Sip{\H o}cz}, B.~M., {et~al.}
    2018, \aj, 156, 123

\bibitem[{{Battaglia} {et~al.}(2011){Battaglia}, {Tolstoy}, {Helmi}, {Irwin},
    {Parisi}, {Hill}, \& {Jablonka}}]{Battaglia11}
{Battaglia}, G., {Tolstoy}, E., {Helmi}, A., {et~al.} 2011, \mnras, 411, 1013

\bibitem[{{Battaglia} {et~al.}(2006){Battaglia}, {Tolstoy}, {Helmi},
    {et~al.}}]{Battaglia06}
---. 2006, \aap, 459, 423

\bibitem[{{Baumgardt} {et~al.}(2019){Baumgardt}, {Hilker}, {Sollima}, \&
    {Bellini}}]{Baumgardt19}
{Baumgardt}, H., {Hilker}, M., {Sollima}, A., \& {Bellini}, A. 2019, \mnras,
    482, 5138

\bibitem[{{Behroozi} {et~al.}(2019){Behroozi}, {Wechsler}, {Hearin}, \&
    {Conroy}}]{Behroozi19}
{Behroozi}, P., {Wechsler}, R.~H., {Hearin}, A.~P., \& {Conroy}, C. 2019,
    \mnras, 488, 3143

\bibitem[{{Bell} {et~al.}(2008){Bell}, {Zucker}, {Belokurov}, {Sharma},
    {Johnston}, {Bullock}, {Hogg}, {Jahnke}, {de Jong}, {Beers}, {Evans},
    {Grebel}, {Ivezi{\'c}}, {Koposov}, {Rix}, {Schneider}, {Steinmetz}, \&
    {Zolotov}}]{Bell08}
{Bell}, E.~F., {Zucker}, D.~B., {Belokurov}, V., {et~al.} 2008, \apj, 680, 295

\bibitem[{{Bellazzini} {et~al.}(2003){Bellazzini}, {Ferraro}, \&
    {Ibata}}]{Bellazzini03}
{Bellazzini}, M., {Ferraro}, F.~R., \& {Ibata}, R. 2003, \aj, 125, 188

\bibitem[{{Bellazzini} {et~al.}(2020){Bellazzini}, {Ibata}, {Malhan}, {Martin},
    {Famaey}, \& {Thomas}}]{Bellazzini20}
{Bellazzini}, M., {Ibata}, R., {Malhan}, K., {et~al.} 2020, \aap, 636, A107

\bibitem[{{Bellazzini} {et~al.}(2006){Bellazzini}, {Newberg}, {Correnti},
    {Ferraro}, \& {Monaco}}]{Bellazzini06}
{Bellazzini}, M., {Newberg}, H.~J., {Correnti}, M., {Ferraro}, F.~R., \&
    {Monaco}, L. 2006, \aap, 457, L21

\bibitem[{{Belokurov} {et~al.}(2006){Belokurov}, {Zucker}, {Evans}, {Gilmore},
    {Vidrih}, {Bramich}, {Newberg}, {Wyse}, {Irwin}, {Fellhauer}, {Hewett},
    {Walton}, {Wilkinson}, {Cole}, {Yanny}, {Rockosi}, {Beers}, {Bell},
    {Brinkmann}, {Ivezi{\'c}}, \& {Lupton}}]{Belokurov06}
{Belokurov}, V., {Zucker}, D.~B., {Evans}, N.~W., {et~al.} 2006, \apjl, 642,
    L137

\bibitem[{{Belokurov} {et~al.}(2014){Belokurov}, {Koposov}, {Evans},
    {Pe{\~n}arrubia}, {Irwin}, {Smith}, {Lewis}, {Gieles}, {Wilkinson},
    {Gilmore}, {Olszewski}, \& {Niederste-Ostholt}}]{Belokurov14}
{Belokurov}, V., {Koposov}, S.~E., {Evans}, N.~W., {et~al.} 2014, \mnras, 437,
    116

\bibitem[{{Ben{\'\i}tez-Llambay} {et~al.}(2016){Ben{\'\i}tez-Llambay},
    {Navarro}, {Abadi}, {Gottl{\"o}ber}, {Yepes}, {Hoffman}, \&
    {Steinmetz}}]{BenitezLlamblay16}
{Ben{\'\i}tez-Llambay}, A., {Navarro}, J.~F., {Abadi}, M.~G., {et~al.} 2016,
    \mnras, 456, 1185

\bibitem[{{Borissova} {et~al.}(2006){Borissova}, {Minniti}, {Rejkuba}, \&
    {Alves}}]{Borissova06}
{Borissova}, J., {Minniti}, D., {Rejkuba}, M., \& {Alves}, D. 2006, \aap, 460,
    459

\bibitem[{{Bovy}(2015)}]{bovy15}
{Bovy}, J. 2015, \apjs, 216, 29

\bibitem[{{Bullock} \& {Johnston}(2005)}]{Bullock05}
{Bullock}, J.~S., \& {Johnston}, K.~V. 2005, \apj, 635, 931

\bibitem[{{Cargile} {et~al.}(2019){Cargile}, {Conroy}, {Johnson}, {Ting},
    {Bonaca}, \& {Dotter}}]{Cargile19}
{Cargile}, P.~A., {Conroy}, C., {Johnson}, B.~D., {et~al.} 2019, arXiv
    e-prints, arXiv:1907.07690

\bibitem[{{Carlin} {et~al.}(2012){Carlin}, {Majewski}, {Casetti-Dinescu},
    {Law}, {Girard}, \& {Patterson}}]{Carlin12}
{Carlin}, J.~L., {Majewski}, S.~R., {Casetti-Dinescu}, D.~I., {et~al.} 2012,
    \apj, 744, 25

\bibitem[{{Carlin} {et~al.}(2018){Carlin}, {Sheffield}, {Cunha}, \&
    {Smith}}]{Carlin18}
{Carlin}, J.~L., {Sheffield}, A.~A., {Cunha}, K., \& {Smith}, V.~V. 2018,
    \apjl, 859, L10

\bibitem[{{Chou} {et~al.}(2007){Chou}, {Majewski}, {Cunha}, {et~al.}}]{Chou07}
{Chou}, M.-Y., {Majewski}, S.~R., {Cunha}, K., {et~al.} 2007, \apj, 670, 346

\bibitem[{{Cioni}(2009)}]{Cioni09}
{Cioni}, M. R.~L. 2009, \aap, 506, 1137

\bibitem[{{Coleman} {et~al.}(2004){Coleman}, {Da Costa}, {Bland-Hawthorn},
    {Mart{\'\i}nez-Delgado}, {Freeman}, \& {Malin}}]{Coleman04}
{Coleman}, M., {Da Costa}, G.~S., {Bland-Hawthorn}, J., {et~al.} 2004, \aj,
    127, 832

\bibitem[{{Conroy} {et~al.}(2019{\natexlab{a}}){Conroy}, {Naidu}, {Zaritsky},
    {Bonaca}, {Cargile}, {Johnson}, \& {Caldwell}}]{Conroy19b}
{Conroy}, C., {Naidu}, R.~P., {Zaritsky}, D., {et~al.} 2019{\natexlab{a}},
    \apj, 887, 237

\bibitem[{{Conroy} {et~al.}(2019{\natexlab{b}}){Conroy}, {Bonaca}, {Cargile},
    {Johnson}, {Caldwell}, {Naidu}, {Zaritsky}, {Fabricant}, {Moran}, {Rhee},
    {Szentgyorgyi}, {Berlind}, {Calkins}, {Kattner}, \& {Ly}}]{Conroy19a}
{Conroy}, C., {Bonaca}, A., {Cargile}, P., {et~al.} 2019{\natexlab{b}}, \apj,
    883, 107

\bibitem[{{Cooper} {et~al.}(2010){Cooper}, {Cole}, {Frenk}, {White}, {Helly},
    {Benson}, {De Lucia}, {Helmi}, {Jenkins}, {Navarro}, {Springel}, \&
    {Wang}}]{Cooper10}
{Cooper}, A.~P., {Cole}, S., {Frenk}, C.~S., {et~al.} 2010, \mnras, 406, 744

\bibitem[{{Deason} {et~al.}(2014){Deason}, {Wetzel}, \&
    {Garrison-Kimmel}}]{Deason14}
{Deason}, A., {Wetzel}, A., \& {Garrison-Kimmel}, S. 2014, \apj, 794, 115

\bibitem[{{Dierickx} \& {Loeb}(2017)}]{DL17}
{Dierickx}, M. I.~P., \& {Loeb}, A. 2017, \apj, 836, 92

\bibitem[{{El-Badry} {et~al.}(2016){El-Badry}, {Wetzel}, {Geha}, {Hopkins},
    {Kere{\v{s}}}, {Chan}, \& {Faucher-Gigu{\`e}re}}]{El-Badry16}
{El-Badry}, K., {Wetzel}, A., {Geha}, M., {et~al.} 2016, \apj, 820, 131

\bibitem[{{Fardal} {et~al.}(2019){Fardal}, {van der Marel}, {Law}, {Sohn},
    {Sesar}, {Hernitschek}, \& {Rix}}]{Fardal19}
{Fardal}, M.~A., {van der Marel}, R.~P., {Law}, D.~R., {et~al.} 2019, \mnras,
    483, 4724

\bibitem[{{Fitts} {et~al.}(2018){Fitts}, {Boylan-Kolchin}, {Bullock},
    {et~al.}}]{Fitts18}
{Fitts}, A., {Boylan-Kolchin}, M., {Bullock}, J.~S., {et~al.} 2018, \mnras,
    479, 319

\bibitem[{{Font} {et~al.}(2001){Font}, {Navarro}, {Stadel}, \&
    {Quinn}}]{Font01}
{Font}, A.~S., {Navarro}, J.~F., {Stadel}, J., \& {Quinn}, T. 2001, \apjl, 563,
    L1

\bibitem[{{Fritz} {et~al.}(2018){Fritz}, {Battaglia}, {Pawlowski},
    {Kallivayalil}, {van der Marel}, {Sohn}, {Brook}, \& {Besla}}]{Fritz18}
{Fritz}, T.~K., {Battaglia}, G., {Pawlowski}, M.~S., {et~al.} 2018, \aap, 619,
    A103

\bibitem[{{Gaia Collaboration} {et~al.}(2018){Gaia Collaboration}, {Brown},
    {Vallenari}, {Prusti}, {et~al.}}]{GaiaDR2}
{Gaia Collaboration}, {Brown}, A.~G.~A., {Vallenari}, A., {Prusti}, T.,
    {et~al.} 2018, \aap, 616, A1

\bibitem[{{Genina} {et~al.}(2019){Genina}, {Frenk}, {Ben{\'\i}tez-Llambay},
    {Cole}, {Navarro}, {Oman}, \& {Fattahi}}]{Genina19}
{Genina}, A., {Frenk}, C.~S., {Ben{\'\i}tez-Llambay}, A.~r., {et~al.} 2019,
    \mnras, 488, 2312

\bibitem[{{Gibbons} {et~al.}(2014){Gibbons}, {Belokurov}, \&
    {Evans}}]{Gibbons14}
{Gibbons}, S.~L.~J., {Belokurov}, V., \& {Evans}, N.~W. 2014, \mnras, 445, 3788

\bibitem[{{Gibbons} {et~al.}(2017){Gibbons}, {Belokurov}, \&
    {Evans}}]{Gibbons17}
---. 2017, \mnras, 464, 794

\bibitem[{{Harris}(1996)}]{Harris96}
{Harris}, W.~E. 1996, \aj, 112, 1487

\bibitem[{{Harris} {et~al.}(2017){Harris}, {Blakeslee}, \& {Harris}}]{Harris17}
{Harris}, W.~E., {Blakeslee}, J.~P., \& {Harris}, G. L.~H. 2017, \apj, 836, 67

\bibitem[{{Hasselquist} {et~al.}(2019){Hasselquist}, {Carlin}, {Holtzman},
    {et~al.}}]{Hasselquist19}
{Hasselquist}, S., {Carlin}, J.~L., {Holtzman}, J.~A., {et~al.} 2019, \apj,
    872, 58

\bibitem[{{Hayes} {et~al.}(2020){Hayes}, {Majewski}, {Hasselquist}, {Anguiano},
    {Shetrone}, {Law}, {Schiavon}, {Cunha}, {Smith}, {Beaton}, {Price-Whelan},
    {Allende Prieto}, {Battaglia}, {Bizyaev}, {Brownstein}, {Cohen},
    {Frinchaboy}, {Garc{\'\i}a-Hern{\'a}ndez}, {Lacerna}, {Lane},
    {M{\'e}sz{\'a}ros}, {Bidin}, {M{\~{u}}noz}, {Nidever}, {Oravetz}, {Oravetz},
    {Pan}, {Roman-Lopes}, {Sobeck}, \& {Stringfellow}}]{Hayes20}
{Hayes}, C.~R., {Majewski}, S.~R., {Hasselquist}, S., {et~al.} 2020, \apj, 889,
    63

\bibitem[{{Helmi} \& {White}(2001)}]{Helmi01}
{Helmi}, A., \& {White}, S. D.~M. 2001, \mnras, 323, 529

\bibitem[{{Hernitschek} {et~al.}(2017){Hernitschek}, {Sesar}, {Rix},
    {Belokurov}, {Martinez-Delgado}, {Martin}, {Kaiser}, {Hodapp}, {Chambers},
    {Wainscoat}, {Magnier}, {Kudritzki}, {Metcalfe}, \& {Draper}}]{Hernitschek17}
{Hernitschek}, N., {Sesar}, B., {Rix}, H.-W., {et~al.} 2017, \apj, 850, 96

\bibitem[{{Hernquist} \& {Mihos}(1995)}]{Hernquist95}
{Hernquist}, L., \& {Mihos}, J.~C. 1995, \apj, 448, 41

\bibitem[{{Hudson} {et~al.}(2014){Hudson}, {Harris}, \& {Harris}}]{Hudson14}
{Hudson}, M.~J., {Harris}, G.~L., \& {Harris}, W.~E. 2014, \apjl, 787, L5

\bibitem[Hunter(2007)]{matplotlib} Hunter, J.~D.\ 2007, Computing in Science and Engineering, 9, 90

\bibitem[{{Ibata} {et~al.}(2020){Ibata}, {Bellazzini}, {Thomas}, {Malhan},
    {Martin}, {Famaey}, \& {Siebert}}]{Ibata20}
{Ibata}, R., {Bellazzini}, M., {Thomas}, G., {et~al.} 2020, \apjl, 891, L19

\bibitem[{{Ibata} {et~al.}(2001){Ibata}, {Irwin}, {Lewis}, \&
    {Stolte}}]{Ibata01}
{Ibata}, R., {Irwin}, M., {Lewis}, G.~F., \& {Stolte}, A. 2001, \apjl, 547,
    L133

\bibitem[{{Ibata} {et~al.}(1994){Ibata}, {Gilmore}, \& {Irwin}}]{Ibata94}
{Ibata}, R.~A., {Gilmore}, G., \& {Irwin}, M.~J. 1994, \nat, 370, 194

\bibitem[{{Ibata} {et~al.}(1995){Ibata}, {Gilmore}, \& {Irwin}}]{Ibata95}
---. 1995, \mnras, 277, 781

\bibitem[{{Ibata} {et~al.}(1997){Ibata}, {Wyse}, {Gilmore}, {Irwin}, \&
    {Suntzeff}}]{Ibata97}
{Ibata}, R.~A., {Wyse}, R. F.~G., {Gilmore}, G., {Irwin}, M.~J., \& {Suntzeff},
    N.~B. 1997, \aj, 113, 634

\bibitem[{{Jiang} \& {Binney}(2000)}]{Jiang00}
{Jiang}, I.-G., \& {Binney}, J. 2000, \mnras, 314, 468

\bibitem[{{Kado-Fong} {et~al.}(2020){Kado-Fong}, {Greene}, {Huang}, {Beaton},
  {Goulding}, \& {Komiyama}}]{Kado-Fong20}
{Kado-Fong}, E., {Greene}, J.~E., {Huang}, S., {et~al.} 2020, arXiv e-prints,
  arXiv:2007.10349

\bibitem[{{Kawata} {et~al.}(2006){Kawata}, {Arimoto}, {Cen}, \&
    {Gibson}}]{Kawata06}
{Kawata}, D., {Arimoto}, N., {Cen}, R., \& {Gibson}, B.~K. 2006, \apj, 641, 785

\bibitem[{{Kazantzidis} {et~al.}(2008){Kazantzidis}, {Bullock}, {Zentner},
    {Kravtsov}, \& {Moustakas}}]{Kazantzidis08}
{Kazantzidis}, S., {Bullock}, J.~S., {Zentner}, A.~R., {Kravtsov}, A.~V., \&
    {Moustakas}, L.~A. 2008, \apj, 688, 254

\bibitem[{{Kirby} {et~al.}(2013){Kirby}, {Cohen}, {Guhathakurta}, {Cheng},
    {Bullock}, \& {Gallazzi}}]{Kirby13}
{Kirby}, E.~N., {Cohen}, J.~G., {Guhathakurta}, P., {et~al.} 2013, \apj, 779,
    102

\bibitem[{{Kruijssen} {et~al.}(2020){Kruijssen}, {Pfeffer}, {Chevance},
    {Bonaca}, {Trujillo-Gomez}, {Bastian}, {Reina-Campos}, {Crain}, \&
    {Hughes}}]{Kruijssen20}
{Kruijssen}, J.~M.~D., {Pfeffer}, J.~L., {Chevance}, M., {et~al.} 2020, arXiv
    e-prints, arXiv:2003.01119

\bibitem[{{Laporte} {et~al.}(2018){Laporte}, {Johnston}, {G{\'o}mez},
    {Garavito-Camargo}, \& {Besla}}]{Laporte18}
{Laporte}, C. F.~P., {Johnston}, K.~V., {G{\'o}mez}, F.~A., {Garavito-Camargo},
    N., \& {Besla}, G. 2018, \mnras, 481, 286

\bibitem[{{Law} \& {Majewski}(2010{\natexlab{a}})}]{LM10}
{Law}, D.~R., \& {Majewski}, S.~R. 2010{\natexlab{a}}, \apj, 714, 229

\bibitem[{{Law} \& {Majewski}(2010{\natexlab{b}})}]{Law10b}
---. 2010{\natexlab{b}}, \apj, 718, 1128

\bibitem[{{Li} {et~al.}(2019){Li}, {Liu}, {Xue}, {Zhong}, {Weiss}, {Carlin}, \&
    {Tian}}]{Li19}
{Li}, J., {Liu}, C., {Xue}, X., {et~al.} 2019, \apj, 874, 138

\bibitem[{{{\L}okas} {et~al.}(2010){{\L}okas}, {Kazantzidis}, {Majewski},
    {Law}, {Mayer}, \& {Frinchaboy}}]{Lokas10}
{{\L}okas}, E.~L., {Kazantzidis}, S., {Majewski}, S.~R., {et~al.} 2010, \apj,
    725, 1516

\bibitem[{{Majewski} {et~al.}(1999){Majewski}, {Siegel}, {Kunkel}, {Reid},
    {Johnston}, {Thompson}, {Land olt}, \& {Palma}}]{Majewski99}
{Majewski}, S.~R., {Siegel}, M.~H., {Kunkel}, W.~E., {et~al.} 1999, \aj, 118,
    1709

\bibitem[{{Majewski} {et~al.}(2003){Majewski}, {Skrutskie}, {Weinberg}, \&
    {Ostheimer}}]{Majewski03}
{Majewski}, S.~R., {Skrutskie}, M.~F., {Weinberg}, M.~D., \& {Ostheimer}, J.~C.
    2003, \apj, 599, 1082

\bibitem[{{Majewski} {et~al.}(2004){Majewski}, {Kunkel}, {Law}, {Patterson},
    {Polak}, {Rocha-Pinto}, {Crane}, {Frinchaboy}, {Hummels}, {Johnston}, {Rhee},
    {Skrutskie}, \& {Weinberg}}]{Majewski04}
{Majewski}, S.~R., {Kunkel}, W.~E., {Law}, D.~R., {et~al.} 2004, \aj, 128, 245

\bibitem[{{Massari} {et~al.}(2019){Massari}, {Koppelman}, \&
    {Helmi}}]{Massari19}
{Massari}, D., {Koppelman}, H.~H., \& {Helmi}, A. 2019, \aap, 630, L4

\bibitem[{{Mateo} {et~al.}(1996){Mateo}, {Mirabal}, {Udalski}, {Szymanski},
    {Kaluzny}, {Kubiak}, {Krzeminski}, \& {Stanek}}]{Mateo96}
{Mateo}, M., {Mirabal}, N., {Udalski}, A., {et~al.} 1996, \apjl, 458, L13

\bibitem[{{Mateo} {et~al.}(1998){Mateo}, {Olszewski}, \& {Morrison}}]{Mateo98}
{Mateo}, M., {Olszewski}, E.~W., \& {Morrison}, H.~L. 1998, \apjl, 508, L55

\bibitem[{{Monachesi} {et~al.}(2019){Monachesi}, {G{\'o}mez}, {Grand },
    {Simpson}, {Kauffmann}, {Bustamante}, {Marinacci}, {Pakmor}, {Springel},
    {Frenk}, {White}, \& {Tissera}}]{Monachesi19}
{Monachesi}, A., {G{\'o}mez}, F.~A., {Grand }, R. J.~J., {et~al.} 2019, \mnras,
    485, 2589

\bibitem[{{Monaco} {et~al.}(2007){Monaco}, {Bellazzini}, {Bonifacio},
    {Buzzoni}, {Ferraro}, {Marconi}, {Sbordone}, \& {Zaggia}}]{Monaco07}
{Monaco}, L., {Bellazzini}, M., {Bonifacio}, P., {et~al.} 2007, \aap, 464, 201

\bibitem[{{Moreno} {et~al.}(2015){Moreno}, {Torrey}, {Ellison}, {Patton},
    {Bluck}, {Bansal}, \& {Hernquist}}]{Moreno15}
{Moreno}, J., {Torrey}, P., {Ellison}, S.~L., {et~al.} 2015, \mnras, 448, 1107

\bibitem[{{Myeong} {et~al.}(2019){Myeong}, {Vasiliev}, {Iorio}, {Evans}, \&
    {Belokurov}}]{Myeong19}
{Myeong}, G.~C., {Vasiliev}, E., {Iorio}, G., {Evans}, N.~W., \& {Belokurov},
    V. 2019, \mnras, 488, 1235

\bibitem[{{Naidu} {et~al.}(2020)}]{Naidu20}
Naidu, R.~P., Conroy, C., Bonaca, A., et al.\ 2020, arXiv:2006.08625

\bibitem[{{Newberg} {et~al.}(2003){Newberg}, {Yanny}, {Grebel}, {Hennessy},
    {Ivezi{\'c}}, {Martinez-Delgado}, {Odenkirchen}, {Rix}, {Brinkmann}, {Lamb},
    {Schneider}, \& {York}}]{Newberg03}
{Newberg}, H.~J., {Yanny}, B., {Grebel}, E.~K., {et~al.} 2003, \apjl, 596, L191

\bibitem[{{Niederste-Ostholt} {et~al.}(2010){Niederste-Ostholt}, {Belokurov},
    {Evans}, \& {Pe{\~n}arrubia}}]{Niederste-Ostholt10}
{Niederste-Ostholt}, M., {Belokurov}, V., {Evans}, N.~W., \& {Pe{\~n}arrubia},
    J. 2010, \apj, 712, 516

\bibitem[{{Oliphant}(2006)}]{numpy}
{Oliphant}, T.~E. 2006, {A guide to NumPy} (Trelgol Publishing USA)

\bibitem[{{Pace} {et~al.}(2020){Pace}, {Kaplinghat}, {Kirby}, {Simon},
    {Tollerud}, {Mu{\~n}oz}, {C{\^o}t{\'e}}, {Djorgovski}, \& {Geha}}]{Pace20}
{Pace}, A.~B., {Kaplinghat}, M., {Kirby}, E., {et~al.} 2020, \mnras,
    arXiv:2002.09503

\bibitem[Perez \& Granger(2007)]{ipython} Perez, F. \& Granger, B.~E.\ 2007, Computing in Science and Engineering, 9, 21

\bibitem[{{Plummer}(1911)}]{Plummer1911}
{Plummer}, H.~C. 1911, \mnras, 71, 460

\bibitem[{Price-Whelan {et~al.}(2017)Price-Whelan, Sipocz, Major, \&
    Oh}]{gala2}
Price-Whelan, A., Sipocz, B., Major, S., \& Oh, S. 2017, adrn/gala: v0.2.1, , ,
    doi:10.5281/zenodo.833339.
\newblock \url{https://doi.org/10.5281/zenodo.833339}

\bibitem[{Price-Whelan(2017)}]{gala1}
Price-Whelan, A.~M. 2017, The Journal of Open Source Software, 2,
    doi:10.21105/joss.00388.
\newblock \url{https://doi.org/10.21105\%2Fjoss.00388}

\bibitem[{{Purcell} {et~al.}(2011){Purcell}, {Bullock}, {Tollerud}, {Rocha}, \&
    {Chakrabarti}}]{Purcell11}
{Purcell}, C.~W., {Bullock}, J.~S., {Tollerud}, E.~J., {Rocha}, M., \&
    {Chakrabarti}, S. 2011, \nat, 477, 301

\bibitem[{{Purcell} {et~al.}(2007){Purcell}, {Bullock}, \&
  {Zentner}}]{Purcell07}
{Purcell}, C.~W., {Bullock}, J.~S., \& {Zentner}, A.~R. 2007, \apj, 666, 20

\bibitem[{{Quinn} \& {Goodman}(1986)}]{Quinn86}
{Quinn}, P.~J., \& {Goodman}, J. 1986, \apj, 309, 472

\bibitem[{{Quinn} {et~al.}(1993){Quinn}, {Hernquist}, \& {Fullagar}}]{Quinn93}
{Quinn}, P.~J., {Hernquist}, L., \& {Fullagar}, D.~P. 1993, \apj, 403, 74

\bibitem[{{Ramos} {et~al.}(2020){Ramos}, {Mateu}, {Antoja}, {Helmi},
    {Castro-Ginard}, \& {Balbinot}}]{Ramos20}
{Ramos}, P., {Mateu}, C., {Antoja}, T., {et~al.} 2020, arXiv e-prints,
    arXiv:2002.11142

\bibitem[{{Revaz} \& {Jablonka}(2018)}]{Revaz18}
{Revaz}, Y., \& {Jablonka}, P. 2018, \aap, 616, A96

\bibitem[{{Rix} \& {Bovy}(2013)}]{Rix13}
{Rix}, H.-W., \& {Bovy}, J. 2013, \aapr, 21, 61

\bibitem[{{Ruiz-Lara} {et~al.}(2020){Ruiz-Lara}, {Gallart}, {Bernard}, \&
    {Cassisi}}]{Ruiz-Lara20}
{Ruiz-Lara}, T., {Gallart}, C., {Bernard}, E.~J., \& {Cassisi}, S. 2020, arXiv
    e-prints, arXiv:2003.12577

\bibitem[{{Sesar} {et~al.}(2017){Sesar}, {Hernitschek}, {Dierickx}, {Fardal},
    \& {Rix}}]{Sesar17}
{Sesar}, B., {Hernitschek}, N., {Dierickx}, M. I.~P., {Fardal}, M.~A., \&
    {Rix}, H.-W. 2017, \apjl, 844, L4

\bibitem[{{Speagle}(2020)}]{dynesty} Speagle, J.~S.\ 2020, \mnras, 493, 3132

\bibitem[{{Skilling}(2004)}]{Skilling04} Skilling, J.\ 2004,
   American Institute of Physics Conference Series, 735, 395

\bibitem[{{Stanimirovi{\'c}} {et~al.}(2004){Stanimirovi{\'c}},
    {Staveley-Smith}, \& {Jones}}]{Stanimirovic04}
{Stanimirovi{\'c}}, S., {Staveley-Smith}, L., \& {Jones}, P.~A. 2004, \apj,
    604, 176

\bibitem[{{Stinson} {et~al.}(2009){Stinson}, {Dalcanton}, {Quinn}, {Gogarten},
    {Kaufmann}, \& {Wadsley}}]{Stinson09}
  {Stinson}, G.~S., {Dalcanton}, J.~J., {Quinn}, T., {et~al.} 2009, \mnras, 395,
    1455

\bibitem[{{Tolstoy} {et~al.}(2004){Tolstoy}, {Irwin}, {Helmi},
    {et~al.}}]{Tolstoy04}
{Tolstoy}, E., {Irwin}, M.~J., {Helmi}, A., {et~al.} 2004, \apjl, 617, L119

\bibitem[{{Totten} \& {Irwin}(1998)}]{Totten98}
{Totten}, E.~J., \& {Irwin}, M.~J. 1998, \mnras, 294, 1

\bibitem[{{van der Marel} {et~al.}(2002){van der Marel}, {Alves}, {Hardy}, \&
    {Suntzeff}}]{vanderMarel02}
{van der Marel}, R.~P., {Alves}, D.~R., {Hardy}, E., \& {Suntzeff}, N.~B. 2002,
    \aj, 124, 2639

\bibitem[{{Velazquez} \& {White}(1999)}]{Velazquez99}
{Velazquez}, H., \& {White}, S. D.~M. 1999, \mnras, 304, 254

\bibitem[Virtanen et al.(2020)]{scipy} Virtanen, P., Gommers, R., Oliphant, T.~E., et al.\ 2020, Nature Methods, 17, 261

\bibitem[{{Yang} {et~al.}(2019){Yang}, {Xue}, {Li}, {Liu}, {Zhang}, {Rix},
    {Zhang}, {Zhao}, {Tian}, {Zhong}, {Xing}, {Wu}, {Li}, {Carlin}, \&
    {Chang}}]{Yang19}
{Yang}, C., {Xue}, X.-X., {Li}, J., {et~al.} 2019, \apj, 886, 154

\bibitem[{{Zolotov} {et~al.}(2009){Zolotov}, {Willman}, {Brooks}, {Governato},
    {Brook}, {Hogg}, {Quinn}, \& {Stinson}}]{Zolotov09}
{Zolotov}, A., {Willman}, B., {Brooks}, A.~M., {et~al.} 2009, \apj, 702, 1058

\end{thebibliography}
\end{document}